\documentclass[11pt, a4paper]{article}
\usepackage{jheppub}
\usepackage{ amssymb }
\usepackage{amsfonts}
\usepackage{amsthm}
\usepackage{braket}
\usepackage{graphicx}
\usepackage{color}
\usepackage{dsfont}
\usepackage{verbatim}
\usepackage{amscd}
\usepackage{multirow}
\usepackage{subfigure}
\usepackage{tikz}
\usepackage{lipsum}  
\usepackage{kantlipsum}  
\usepackage{comment}
\usepackage{parskip}
\usepackage[export]{adjustbox}
\usepackage{upgreek}
\usepackage{ bbold }
\usepackage{ dsfont }
\usepackage{accents}

\usepackage{commath}
\usepackage{slashed}
\usepackage{braket}

\newcommand*{\GR}[1]{\textcolor{purple}{GR :{#1}}}

\newcommand*{\bfn}{{ \bf n }}
\newcommand*{\bfm}{{ \bf m }}
\newcommand*{\bfei}{{ \bf e_i }}
\newcommand*{\bfej}{{ \bf e_j }}
\newcommand*{\bfex}{{ \bf e_x }}
\newcommand*{\bfey}{{ \bf e_y }}

\newcommand*{\one}{ \mathds{1} }

\newcommand*{\Uph}{\mathds{U}}
\newcommand*{\Dph}{\mathds{D}}

\usepackage{microtype}
\begin{document}

\title{Gauging a superposition of fermionic Gaussian projected entangled pair states to get lattice gauge theory eigenstates}

\begin{abstract}{
    Gauged fermionic projected entangled pair states (GFPEPS) and their Gaussian counterpart (GGFPEPS) are a novel type of lattice gauge theory Ansatz state that combine ideas from the Monte Carlo and tensor network communities. In particular, computation of observables for such states boils down to a Monte Carlo integration over possible gauge field configurations that have probabilities dictated by a fermionic tensor network contraction that accounts for the matter in that background configuration. Crucially, this probability distribution is positive definite and real so that there is no sign problem.
    
    When the underlying PEPS is Gaussian, tensor network contraction can be done efficiently, and in this scenario the Ansatz has been tested well numerically. In this work we propose to gauge superpositions of Gaussian PEPS and demonstrate that one can still efficiently compute observables when few Gaussians are in the superposition. As we will argue, the latter is exactly the case for bound states on top of the strongly interacting LGT vacuum, which makes this Ansatz particularly suitable for that scenario.
  
    As a corollary, we will provide an exact representation of the LGT ground state as a gauged PEPS. 
}\end{abstract}

\author[a, b]{Gertian Roose}
\author[a, b]{and Erez Zohar}
\affiliation[a]{Racah institute of Physics, The Hebrew University of Jerusalem, Givat Ram, Jerusalem 91904, Israel}
\affiliation[b]{School of Physics and Astronomy, Tel Aviv University, Tel Aviv 6997801, Israel}
\emailAdd{Gertian.Roose@mail.huji.ac.il} 

\maketitle	
\flushbottom
\section{Introduction}
Since their inception in the 1950s, gauge theories have formed the cornerstone for our understanding of nature. In the context of condensed matter physics they are often used as effective low energy descriptions of spin liquids \cite{Fradkin_Field_Theories_of_Condensed_Matter_Physics,Savary_Quantum_spin_liquids}. In high energy physics they have been applied to model the interactions between particles in the standard model with astonishing accuracy, in this context the most famous example is quantum chromodynamics (QCD) which is a non-Abelian gauge theory that describes the interactions between quarks and gluons \cite{Peskin_Introduction_to_QFT}. 

Despite these formidable successes, our understanding of gauge theories is far from complete due to the fact that gauge theories have very rich phase diagrams. In particular, for large couplings the infrared degrees of freedom are colorless hadrons and mesons which cannot be described through perturbation theory in terms of quarks and gluons \cite{Gross_Assymptotically_free_gauge_theories}. Similarly, at low temperature and high fermion density it is expected that the emergent IR degree of freedom are color-Cooper pairs instead of some perturbative correction to free quarks and gluons \cite{Reddy_Novel_Phases_at_High_Density_and_their_Roles_in_the_Structure_and_Evolution_of_Neutron_Stars,RAJAGOPAL_THE_CONDENSED_MATTER_PHYSICS_OF_QCD}. 

To combat this shortcoming of perturbation theory, Wilson proposed (1974) to discretize the Euclidean (i.e. Wick rotated) theory on a lattice where it can be solved numerically \cite{Wilson_Confinement_of_quarks,Kogut_introduction_to_lattice_gauge_theory}. Historically the leading method to do this was Monte Carlo sampling of the path integral which, when applicable, is extremely powerful \cite{Creutz_Monte_Carlo_study_quantized_SU2, Creutz_Monte_Carlo_LGT}. This is demonstrated by the fact that it has been used to calculate the mass of many hadrons to a few percent accuracy \cite{Aoki_FLAG_Review_2021}. However, this method has two major drawbacks. First, the Wick rotated approach does not allow the direct study of real time evolution. Second, in many interesting scenarios such as the before mentioned high fermion density, the probability distribution over which the Monte Carlo integration is performed becomes non-positive or even complex, leading to the notorious sign problem \cite{Troyer_Computational_Complexity_Fermionic_QMC}.

The Hamiltonian lattice discretization of lattice gauge theory was introduced slightly later by Kogut and Susskind \cite{Kogut_Hamiltonian_formulation_of_lattice_gauge_theories, Susskind_Hamltonian_lattice_fermions}. This formulation has the advantage that time is a real continuous variable so that time evolution is readily available. Furthermore, computation of observables reduces to solving eigenvalue problems which do not suffer from a sign problem. However, as is well known, the Hilbert space of the Hamiltonian theory is exponentially large in the number of lattice sites which makes it impossible to solve the eigenvalue problem exactly. To combat this, one must propose an Ansatz state that parameterizes a subspace of the full Hilbert space and solve the eigenvalue problem in that subspace. 

One such Ansatz is the class of tensor network states (TNS) \cite{White_Density_matrix_formulation_QRG,Fannes_Finitely_correlated_states,Schollwock_the_density_matrix_RG_in_the_age_of_MPS,Cirac_matrix_product_states_and_projected_entangled_pair_states,Jordan_Classical_Simulation_of_2D_Lattice_Systems,Cirac_Renormalization_and_tensor_product_states,Orus_Practical_introduction_to_tensor_networks}. In particular, in 1+1 dimensions matrix product states (MPS) have been applied very successfully to model the ground state and time evolution of lattice gauge theory (LGT) Hamiltonians \cite{Banuls_the_mass_spectrum_of_the_Schwinger_model,Buyens_Matrix_Product_States_for_Gauge_Field_Theories,Rico_Tensor_Networks_for_Lattice_Gauge_Theories,Kuhn_Non_Abelian_string_breaking_phenomena_with_MPS,Banuls_Thermal_evolution_of_the_Schwinger_model_with_MPO,Pichler_Real_Time_Dynamics_in_U1_LGT_with_Tensor_Networks,Buyens_Hamiltonian_simulation_of_the_Schwinger_model,Dalmonte_Lattice_gauge_theory_simulations_in_the_quantum_information_era,Banuls_Simulating_lattice_gauge_theories_within_quantum_technologies,Banuls_Review_on_novel_methods_for_lattice_gauge_theories}. However, exact contraction of the straightforward higher dimensional generalization of MPS, i.e. projected entangled pair states (PEPS), scales exponentially with the system size \cite{Schuch_Computational_Complexity_of_PEPS}. Nevertheless, TNS have been successfully applied to study higher dimensional LGT \cite{Tagliacozzo_Tensor_Networks_for_Lattice_Gauge_Theories_with_Continuous_Groups,Crone_Detecting_Z2_topological_order_from_iPEPS,Robaina_Simulating_2+1D_Z3_LGT_with_iPEPS}. In practice, the origin of the exponential scaling can be traced back to the presence of loops in PEPS \cite{Schuch_Computational_Complexity_of_PEPS}, therefore one way to circumvent this scaling is to consider tree tensor networks (TTN) that are loop free. These have been successfully used to study LGT in up to 3+1 dimensions \cite{Magnifico_lattice_QED_in_3+1D_with_tensor_networks,Montangero_loop_free_tensor_networks_for_high_energy_physics,Cataldi_Simulating_2+1D_SU2_Yang_Mills_lattice_gauge_theory_at_finite_density_with_tensor_networks, Felser_two_dimensional_quantum_link_lattice_quantum_electrodynamics}. Additionally, the need for approximate contraction of higher dimensional tensor networks has led to the development of coarse-graining methods inspired by the renormalization group. Such algorithms have been developed and applied successfully for the Hamiltonian \cite{Tagliacozzo_Entanglement_renormalization_and_gauge_symmetry} (i.e. coarse graining states) and Lagrangian \cite{Meurice_Tensor_lattice_field_theory_for_renormalization_and_quantum_computing} (i.e. coarse graining path integrals) approaches to LGT. Finally, one may suppress the exponential cost by working with Gaussian tensor networks \cite{Shuch_Gaussian_Matrix_Product_States,Mortier_Tensor_Networks_Can_Resolve_Fermi_Surfaces, Kraus_fermionic_projected_entangled_pair_states} that are efficiently contractible in the framework of \cite{Bravyi_Lagrangian_representation_fermionic_linear_optics}. 

One major advantage of all TNS is that they implement global symmetries of the state through local symmetries of the tensors that build the network\cite{Singh_Tensor_network_decompositions_in_the_presence_of_a_global_symmetry}. Furthermore, this idea can be extended beautifully to states with local symmetries for LGT. This is done by using group averaging in \cite{Haegeman_Gauging_Quantum_States_Global_Local_Symmetries} or using so-called gauging tensors in \cite{Zohar_Combining_tensor_networks_with_Monte_Carlo_methods_for_LGT, Zohar_Building_projected_entangled_pair_states_with_local_gauge_symmetry}. Recently, it was proven that all locally symmetric PEPS can be written as a gauged PEPS cfr. \cite{Kull_Classification_of_matrix_product_states_with_local_gauge_symmetry} in 1+1D and \cite{Blanik_internal_structure_of_gauge_invariant_projected_entangled_pair_states} in arbitrary dimension. To allow for efficient contraction of these states one can constrain the underlying PEPS to be Gaussian. The resulting gauged Gaussian Fermionic PEPS (GGFPEPS) that combine Monte Carlo integration over possible gauge field configurations with fermionic Gaussian PEPS states on the fermions have already been used for ground state searches of pure $\mathbb{Z}_2$ and $\mathbb{Z}_3$ LGT \cite{Emonts_variational_monte_carlo_simulation_with_tensor_networks_of_a_pure_z3_gauge_theory,Emonts_finding_ground_state_of_lattice_gauge_theory_with_fermionic_tensor_networks,Zohar_Combining_tensor_networks_with_Monte_Carlo_methods_for_LGT,Kelman_Gauged_Gaussian_projected_entangled_pair_states_A_high_dimensional_tensor_network_formulation_for_lattice_gauge_theories} as well as $\mathbb{Z}_2$ lattice gauge theory with fermions \cite{Kelman_Gauged_Gaussian_projected_entangled_pair_states_A_high_dimensional_tensor_network_formulation_for_lattice_gauge_theories}. 

In the present work, we demonstrate that all lattice gauge theory ground states are exactly represented by gauged PEPS that are non-Gaussian only on the bonds, i.e. gauged PEPS that have purely Gaussian on-site tensors and non-Gaussian bond tensors. Note that this is a stronger result than the one in cfr \cite{Kull_Classification_of_matrix_product_states_with_local_gauge_symmetry}. In proving this we will exactly construct the gauged PEPS which offers valuable physical insight into the construction. Next we will explore the scenario where these non-Gaussian bonds are considered to be superpositions of Gaussian bonds, and we will show that such states can be efficiently contracted when few Gaussian PEPS are superposed. Since the strongly interacting LGT vacuum is purely Gaussian, it is expected that this will be the case when considering isolated bound states such as the hadrons and mesons on top of this strongly interacting vacuum. 

Note that we will not explicitly implement the gauged superposition of Gaussian PEPS numerically in this work. The rationale for this is that it is merely a generalization of the already implemented gauged Gaussian PEPS which so far has proven sufficient for all scenarios studied \cite{Emonts_variational_monte_carlo_simulation_with_tensor_networks_of_a_pure_z3_gauge_theory,Emonts_finding_ground_state_of_lattice_gauge_theory_with_fermionic_tensor_networks,Zohar_Combining_tensor_networks_with_Monte_Carlo_methods_for_LGT,Kelman_Gauged_Gaussian_projected_entangled_pair_states_A_high_dimensional_tensor_network_formulation_for_lattice_gauge_theories, Kelman_Gauged_Gaussian_projected_entangled_pair_states_A_high_dimensional_tensor_network_formulation_for_lattice_gauge_theories}. Only when the purely Gaussian Ansatz reaches its limit we will need to implement the proposed algorithm. 

The outline of the paper is as follows. In section 2 we review lattice gauge theories in D-dimensions. In section 3 we introduce PEPS, demonstrate the gauging procedure and discuss the Gaussian subset of PEPS. In section 4 we posit the gauged superposition of Gaussian PEPS and in section 5 we show how to compute gauge invariant observables for it. In section 6 we provide the exact gauged PEPS representation of the LGT ground state and show that it is indeed sufficient to have only non-Gaussian bonds. 

\section{Foundations of Hamiltonian lattice gauge theory}
Before we delve into lattice gauge theories, let us first look at the more familiar staggered lattice discretization (let us denote the lattice spacing as $a$) for a set of $N$ fermions $\psi_{\bfn \alpha} = \psi^{\text{lattice}}_{\bfn \alpha} = a^{D/2} \psi^{QFT}_\alpha(\bfn)$ where $\bfn \in $ \emph{lattice points} and $\alpha \in 1:N$ \cite{Kogut_Hamiltonian_formulation_of_lattice_gauge_theories,Kogut_introduction_to_lattice_gauge_theory}. The Hamiltonian for this system, using the notations $\sum_\bfn = \sum_{\bfn \in \text{lattice}}$ and $\sum_{<\bfn \bfm>} = \sum_{\substack{\bfn \in \text{lattice}\\ \bfm = \bfn + \bfei \forall \bfei }}$ where $i$ and $\bfei$ respectively represent spatial dimensions and unit vectors in each direction $\bfei$,  is given by: 
\begin{align}
    H_{\text{free fermion}} = \underbrace{\frac{i}{a}\sum_{\langle {\bf n}{\bf m} \rangle} {\eta}_{\bfn \bfm} \  \psi_{\bfn \alpha}^\dagger \psi_{\bfm \alpha} + h.c.}_{H_{\text{hopping}}} + \underbrace{\frac{1}{a}\mu a\sum_{\bfn} \psi_{\bfn \alpha}^\dagger \psi_{\bfn \alpha}}_{H_{\text{chemical}}} + \underbrace{\frac{1}{a}m a \sum_{\bfn} (-1)^\bfn \psi_{\bfn \alpha}^\dagger \psi_{\bfn \alpha}}_{H_{\text{mass}}}
\end{align}
and consists of a hopping term in which $\eta_{\bfn \bfm}$ are some phases implementing a $\pi$-flux around each plaquette, an anti-ferromagnetic mass term and a chemical potential term that favors a finite fermion density\footnote{All choices for $\eta_{\bfn \bfm}$ work as long as they implement the $\pi$-flux. One example is $\eta_{\bfn \bfm} = 1$ on all links except the odd horizontal ones where it is -1. On top of this freedom in $\eta_{\bfn \bfm}$ there is also freedom in the choice of mass term. Indeed, instead of the antiferromagnetic mass term one could choose a dimerizing mass term $\frac{1}{a}\frac{ma}{2}\sum_{<\bfn \bfm>} (-1)^\bfn \psi^\dagger_{\bfn \, \alpha} \psi_{\bfm \, \alpha}$. It can be shown that the choice in $\eta_{\bfn \bfm}$ as well as the choice in mass term can be related to the choice of QFT $\gamma$-matrices before staggering. Furthermore, it can be shown that the choices presented in this footnote most closely capture the topological properties of the free fermion. \cite{Roose_Lattice_regularisation_and_entanglement_structure_of_the_Gross_Neveu_model,Roose_the_chiral_Gross_Neveu_model_on_the_lattice_via_a_Landau_forbidden_phase_transition}}. Note that $\mu$ and $m$ are both dimensionful couplings that directly correspond to the QFT chemical potential and fermion mass respectively.

It is well known that this Hamiltonian has global non-anomalous $SU(N)$ and $U(1)$ symmetries generated by :
\begin{align} 
    \theta_{SU(N)}(g) &= \prod_{\bfn} \exp\del{{i \phi_a(g) \underbrace{\psi_{\bfn \alpha}^\dagger T^{a}_{\alpha \beta} \psi_{\bfn \beta}}_{ Q^a_\bfn}  } } \\
    \theta_{U(1)}(g) &= \prod_{\bfn} \exp\del{{i \phi(g) \psi_{\bfn \alpha}^\dagger \psi_{\bfn \alpha} } }
\end{align}    
with $T^a$ the $N^2-1$ generators of the $SU(N)$ group, usually in the fundamental irrep (irreducible representation), that satisfy the group algebra $[T^a, T^b] = i f_{a b c} T^c $ where $f_{abc}$ are the $SU(N)$ structure constants. For $SU(2)$ in the fundamental $j=1/2$ irrep these are the Pauli matrices over 2. Again for $SU(2)$ in the $j=1$ irrep these are the spin-1 matrices and for $SU(3)$ in the fundamental irrep these will be the Gell-Mann matrices.

As is shown in appendix \ref{appendix:Review on relevant group actions} the action of these group elements on the fermionic operators is given by :
\begin{align}
    \theta_{SU(N)}(g) \ \psi^\dagger_{\bfn\alpha} \ \theta_{SU(N)}(g)^\dagger &= \psi^\dagger_{\bfn \beta} \ \overbrace{\exp\del{i \phi_a(g) T^a_{\beta \alpha} }}^{D_{\beta \alpha}(g)}    \\
    \theta_{U(1)}(g)_{\phantom{U}} \ \psi^\dagger_{\bfn\alpha} \ \theta_{U(1)}(g)^\dagger_{\phantom{U}} &= \psi^\dagger_{\bfn \alpha} \  \exp\del{i \phi(g)} 
\end{align}
where we have identified the Wigner matrix $D_{\beta \alpha}(g)$ that provides a linear right acting regular implementation of the group action. It is easy to see that $[H_\text{hopping}, \theta_{SU(N)}(g)] = [H_\text{hopping}, \theta_{U(1)}(g)] = 0$ which leads to the conservation of color charge and total fermion number respectively. 

Broadly speaking, the goal of (lattice) gauge theories is to make these symmetries local through the introduction of new Hilbert space associated to the links on which flux lines reside. More formally, $\theta(g)$ becomes $\theta_\bfn(g_\bfn)$ which now acts like $\theta_\bfn(g_\bfn) \psi^\dagger_{\bfn \alpha} \theta^\dagger_\bfn(g_\bfn) =\psi^\dagger_{\bfn \beta} D_{\beta \alpha}(g_\bfn) $ on the fermions. For now, let us simply assume the existence of $SU(N)$ charged operators $U^{j;\alpha \beta}_{\bfn \bfm}$ that reside on the link in between nearest neighbors $\bfn$ and $\bfm=\bfn + \bfei$ and transforms as : 
\begin{align}
    \theta_\bfn(g_\bfn)\theta_\bfm(g_\bfm) \ U^{j; \alpha \beta}_{\bfn\bfm} \  \theta_\bfn^\dagger(g_\bfn)\theta_\bfm^\dagger(g_\bfm) = D^j_{\alpha \alpha' }(g^{-1}_{\bfn})\  U^{j; \alpha' \beta'}_{\bfn \bfm} \  D^j_{\beta' \beta}(g_{\bfm})  \label{eq:guess_U}   \ \ .
\end{align}
With those, we can define the gauged hopping Hamiltonian as
\begin{align}
    H_{\text{hopping}} = \frac{i}{a} \sum_{\langle {\bf n} {\bf m} \rangle} \eta_{\bfn \bfm} \psi_{\bfn \alpha}^\dagger U_{\bfn \bfm}^{\alpha\beta}  \psi_{\bfm \beta} + h.c. 
\end{align}
which now commutes with $\theta_\bfn(g_\bfn)$ $\forall \bfn$. Note that we have denoted $U^{\alpha \beta}_{\bfn \bfm}$ for  $U^{j = \text{fundamental}; \alpha \beta}_{\bfn \bfm}$. 

\subsection{The link Hilbert space}
\emph{This section will be an overview of the minimal mathematical machinery required to understand the link Hilbert space. For more detail we refer to section 5.6 of \cite{Nakahara_Geometry_topology_and_physics}.}

Let us focus on a single link between sites $\bfn$ and $\bfm$ and let us note that $U_{\bfn \bfm}^{\alpha \beta}$ (or shorthand $U^{\alpha \beta}$ throughout this subsection) carries distinct charges associated with transformations on $\bfn$ and $\bfm$. Furthermore, \ref{eq:guess_U} reveals that these transformations are respectively left inverse and right regular so that their respective generators $L^a$ and $R^a$ should satisfy (cfr. appendix \ref{appendix:Review on relevant group actions}) : 
\begin{align}
    [  L^a , L^b ] &= -i f_{a b c} L_c  \label{eq:linkalgebra} \\
    [  R^a , R^b ] &= \phantom{-}i f_{a b c} R_c  \\
    [ L^a , R^b ] &= \phantom{-i} 0 
\end{align}
where we have dropped the subscripts $\bfn$ and $\bfm$ to avoid unnecessary cluttering. 

To start building the vector space on which these act, we first define the left and right Cartan subalgebras denoted $\mathbb{h_l}$ and $\mathbb{h_r}$ as a choice of maximal mutually commuting subgroups of $L^a$ and $R^a$ respectively. For $SU(2)$ the left and right Cartan subalgebras are usually chosen to be $L_z$ and $R_z$. For $SU(3)$ they consist of any two mutually commuting $L^a$ and any two mutually commuting $R^a$. Similarly, for $SU(N)$ they will consist of any $N-1$ mutually commuting  $L^a$ and any $N-1$ mutually commuting $R^a$ operators.  

Additionally, we define the left and right Casimir operators $C^i_L$ and $C^i_R$ (respectively constructed from the $L^a$ and $R^a$) as maximal subsets of operators build from the $L^a$ and $R^a$ that commute with each other and all other $L^a$ and $R^a$. For example, when $G=SU(2)$ there is a single left Casimir operator $C^1_L = L^2=\sum_a L^a L^a$ and a single right Casimir operator $R^2$. For $SU(3)$ the situation is a bit more complicated; we first need to define $d_{abc}$ so that $\{ L_a , L_b\} = \frac{1}{3}\delta_{ab} \one + d_{abc} L_c$. With this we get two left Casimir operators $C^1_L = L^2$ and $C^2_L = \sum_{abc} d_{abc} L_a L_b L_c$ and similar for the right side of the link\footnote{In the mathematics literature the number of generators appearing in a Casimir element is named the order of that Casimir operator.}. In the general case of $SU(N)$ there will be $N-1$ Casimir operators on each side of the link.  

It can then be shown that the mutual eigenvectors of $\{C^i_L\}$, $\{C^i_R\}$, $\{L^a \in \mathbb{h_l} \}$ and $\{R^a \in \mathbb{h_l} \}$ form a basis for the link Hilbert space. Furthermore, labeling these vectors by integers $\{j^i_l\}$, $\{j^i_r\}$, $\{n_i\}$ and $\{m_i\}$ that are respectively related to the eigenvalues of $\{C^i_L\}$, $\{C^i_R\}$, $\{L^a \in \mathbb{h_l} \}$ and $\{R^a \in \mathbb{h_l} \}$ and noting that we are only interested in those vectors for which $j^i_l = j^i_r = j \, \forall \, i$ leads to basis vectors $\ket{j n m } =\ket{\{j^i\} \{n^i\}} \otimes \ket{\{j^i\} \{m^i\} }$ with $\{j\} \in irreps$ and $\{n\}$ and $\{m\}$ both in the quantum numbers for that irrep. In shorthand $\{n \} $ and $\{m\}$ $\in QN(j)$. As an example, for $SU(2)$ we have the fundamental irrep $j=1/2$ where both $n$ and $m$ run over $\pm 1/2$, the $T^a$ are the Pauli matrices which are indeed $2\times 2$. Also, for $SU(2)$ we have the $j=1$ irrep where $n$ and $m$ run over $-1,0,1$ and the $T^a$ are the so called spin-1 matrices that are $3 \times 3$. For general $SU(2)$ representations we have : $L^2\ket{j n m} = R^2 \ket{j n m} = j(j+1) \ket{j n m}$, $L_z \ket{j n m } = n \ket{j n m}$ and $R_z \ket{j n m } = m \ket{j n m}$. For $SU(3)$ we have the fundamental irrep $(j^1, j^2)=(1,0)$ and both $n=(n^1, n^2)$ and $m=(m^1, m^2)$ run over $(1,1),(-1,1),(0,2)$; in the same irrep the $T^a$ are the Gell-Mann matrices which are $3\times 3$ matrices. 

Apart from the representation basis, we will also need the so-called group element basis $\ket{g}$. This one is defined through 
\begin{align}
    \ket{g} = \sum_{\substack{j \in irreps\\ n,m \in QN(j)}} \ \sqrt{\frac{\dim(j)}{|G|}} \ D^j_{mn}(g^{-1}) \ket{j n m}  \  \ \forall g \in G
\end{align}
where $|G|$ is the group's order when $G$ is a finite group and the group's volume $\int dg$ when $G$ is a compact Lie group. For the $SU(2)$ example we find $\dim(j) = 2j +1$ and $|G| = 8 \pi^2$. Note how the indices on the Wigner matrix are switched compared to the ket. It can be shown that the representation basis $\ket{j n m}$ and group element basis $\ket{g}$ both constitute normalized bases for the Hilbert space consequently $\braket{g|g'} = \delta(g, \,g')$ and $\braket{j n m | j' n' m'} = \delta_{j j'} \delta_{n n'} \delta_{m m'}$ and the resolution of the identity in both bases is $\one = \int dg \, \ket{g}\bra{g} = \sum_{\substack{j \in irreps\\ n,m \in QN(j)}} \ket{j n m}\bra{j n m} = \one$. Additionally, this is a great time to note that $\bar D^j_{nm}(g) = D^j_{mn}(g^{-1})$ as this is required for the inverse transformation. 
 
Now that we have constructed bases for the link Hilbert space we can start to formally define operators that act on it. For starters, it is easy to see that 
\begin{align}
    L^a = \sum_{\substack{j \in irreps\\ n,n', m \in QN(j)}} \ket{j n m} (T^a)^j_{n' n} \bra{j n' m}
\end{align}
and
\begin{align}
    R^a = \sum_{\substack{j \in irreps\\ n,m, m' \in QN(j)}} \ket{j n m} (T^a)^j_{m m'} \bra{j n m'} 
\end{align}
satisfy the desired commutation relations \ref{eq:linkalgebra}. Exponentiation of these generators leads to group elements :
\begin{align}
    \theta^L(g_L) = \exp\del{- i \phi_a(g_L)\  L^a } \\
    \theta^R(g_R) = \exp\del{+ i \phi_a(g_R) \  R^a } 
\end{align}
and it is straightforward to show that they act as 
\begin{align}
    \theta^L(g_L)\theta^R(g_R) \ket{j m n} &= D^j_{m m'}(g_L^{-1}) \ket{j m' n'} D^j_{n' n}(g_R)  \\
    \theta^L(g_L)\theta^R(g_R) \ket{h} &= \ket{g_L \  h \  g_R^{-1}} 
\end{align}
on the two types of basis vectors. Note that the $\theta^L$ was exponentiated with a minus sign to ensure $\ket{jmn}$ transforms as a left inverse representation rather than a left regular representation.

Now we are finally ready to define
\begin{align}
    U^{j; \alpha \beta} = \int dg \  D^j_{\alpha \beta}(g) \ket{g}\bra{g}  \label{eq:def_U}
\end{align}
which, using the transformation properties of $\Ket{g}$, can be shown to transform as
\begin{align}
    \theta^R(g_R)\theta^L(g_L) \ U^{j; \alpha \beta} \  \theta^{L \dagger}(g_L)\theta^{R\dagger}(g_R) = D^j_{\alpha \alpha'}(g^{-1}_L) \  U^{j; \alpha' \beta'} \ D^j_{\beta' \beta}(g_R)  \label{eq:U_transform}  \ \ .
\end{align}

Taking $\theta^L(g_L)$ close to the identity leads to $[L^a, U^{j; \alpha \beta}] = (T^j)^a_{\alpha \delta} U^j_{\delta \beta}$. Similarly, we can find $[R^a, U^{j; \alpha \beta}] = U^j_{\alpha \delta} (T^j)^a_{\delta \beta}$ so that $U^j_{\alpha \beta}$ is a creation operator for the dimensionless electrical fields $L^a$ and $R^a$ that live on the left and right side of links. 
\footnote{$U$ is related to the QFT connection $A^a_\mu(x)$ through $U^{j;\alpha \beta}_{\bfn \bfm} = \exp(i\  ag_\text{qft}\ (T^a)^j_{\alpha \beta} A^{j \, a}_{\mu = {\bf m - n }}( \frac{\bfm + \bfn}{2} ))$ and similarly, $L^a_{\bfn \bfm} = \frac{a^{D-1}}{g} E(\frac{\bfn+\bfm}{2} ) $. Using these, it can be shown that the commutation relations $[L^a, U^{j; \alpha \beta}] = (T^j_a)_{\alpha \delta} U^j_{\delta \beta}$ imply the better known canonical commutation relations for the color-electric field and 4-potential $[A^a(x), L^b(y)] = i \delta_{ab} \delta(x-y)$ .} 
\footnote{For completeness, and to avoid confusion over mathematical subtleties, we note that \ref{eq:def_U} and \ref{eq:U_transform} imply $ (U^{j;\alpha \beta})^\dagger = \int dg \, D^j_{\beta \alpha}(g^{-1}) \ket{g}\bra{g} $ and $ \theta^R(g_R)\theta^L(g_L) \ (U^{j; \alpha \beta})^\dagger \  \theta^{L \dagger}(g_L)\theta^{R\dagger}(g_R) = D^j_{\alpha' \alpha}(g_L) \  (U^{j; \alpha' \beta'})^\dagger \ D^j_{\beta \beta'}(g^{-1}_R)$. Indeed, this can be easily demonstrated using $D^*_{\alpha \beta}(g) = D_{\beta \alpha}(g^{-1})$.}

\subsection{The local group elements}
Using these link Hilbert space and operators, we are finally ready to define the gauge group elements as
\begin{align}
    \theta_\bfn(g_\bfn) = \theta(g_\bfn) = \exp\del{i \phi_a(g) \ G^a_\bfn } \label{eq:local charge}
\end{align}
with
\begin{align}
    G^a_\bfn = Q^a_\bfn - \sum_{\bf e_i} \underbrace{\del{L^a_{\bfn, \bfn + {\bf e_i}} - R^a_{\bfn - {\bf e_i} ,\bfn }}}_{\substack{\text{discretized divergence } \\ \text{of the color-electric field}}}   
\end{align}
where 
\begin{align}
    Q^a_\bfn &= \Psi^\dagger_{\bfn \alpha} \  (T^a)^{j = j_f}_{\alpha \beta} \ \Psi_{\bfn  \beta} \\
    L^a_{\bfn, \bfn + \bfei} &= \sum_{\substack{j \in irreps\\ n,n', m \in QN(j)}} \ket{j n m}_{\bfn, \bfn + \bfei} (T^a)^j_{n' n} \bra{j n' m}_{\bfn, \bfn + \bfei} \\
    R^a_{\bfn - \bfei, \bfn} &= \sum_{\substack{j \in irreps\\ n,m, m' \in QN(j)}} \ket{j n m}_{\bfn - \bfei, \bfn} (T^a)^j_{m m'} \bra{j n m'}_{\bfn - \bfei, \bfn}  \ \ .
\end{align}
and $j_f$ is the representation of the matter that resides on the vertices. Indeed, with this the fermions and link variables transform as discussed before. 

\subsection{Plaquette and electric field terms}
Since the gauged Hamiltonian now acts on an extended Hilbert space, it is natural to add terms that act purely on this new link Hilbert space \cite{Kogut_Hamiltonian_formulation_of_lattice_gauge_theories}. Guided by the principle that these terms should also respect the local $SU(N)$ symmetry we construct the magnetic energy $H_B = H_\square + h.c.$ where : 
\begin{align}
    H_{\square} &= \frac{1}{a}\frac{4a^{D-3}}{g^2_\text{qft}}\sum_{\substack{ \bf n  \\ {\bf e_1, e_2} } }  \overbrace{U^{\alpha \beta}_{ \bfn, \bfn+\bf{e_1} } U^{\beta \gamma}_{\bf n+e_1, n+e_1+e_2} U^{\alpha \delta\ \dagger}_{\bf n, n+e_2} U^{\delta \gamma\ \dagger}_{\bf n + e_2, n+e_1+e_2}}^{U_{\bf n, e_1, e_2}^{\square}}  \label{eq:H_square}
\end{align}
and $g_\text{qft}$ is the dimension-full QFT coupling constant. This term creates (or annihilates) closed loops of color-electrical flux. Since both $U$ and $U^\dagger$ are diagonal in the gauge field configuration basis it is easy to show that 
\begin{align}
    \braket{\mathcal{G}| U^\square_{\bfn \, \bfei \, \bfej } |\mathcal{G}} &= D_{\alpha \beta}(g_{\bfn \, \bfn+\bfei})   D_{\beta \gamma}(g_{\bfn+\bfei \, \bfn+\bfei+\bfej}) D_{\delta \alpha}(g^{-1}_{\bfn \, \bfn+\bfej}) D_{\gamma \delta}(g^{-1}_{\bfn+\bfej \, \bfn+\bfei+\bfej}) \\
    &=  D_{\alpha \alpha}\del{ g_{\bfn \, \bfn + \bfei} g_{\bfn+\bfei \, \bfn+\bfei+\bfej} g^{-1}_{\bfn + \bfej \, \bfn+\bfei+\bfej} g^{-1}_{\bfn \, \bfn + \bfej} } \\
    &= D^\square_{\bfn \, \bfei \, \bfej} (\mathcal{G}) 
\end{align}
where $\ket{\mathcal{G}} = \otimes_{\braket{\bfn \bfm}} \ket{g_{
    \bfn \bfm
}}$ has been defined as a gauge field configuration on the entire lattice. In similar fashion we define $H_\text{hopping}(\mathcal{G}) = \braket{\mathcal{G}|H_\text{hopping}|\mathcal{G}}$.

Another possible term is :
\begin{align}
   H_{E} &= \frac{1}{a}\frac{g^2_\text{qft}}{2 a^{D-3}}\sum_{ \langle {\bf n}, {\bf m} \rangle  }  J^2_{\bf nm} \label{eq:H_E}
\end{align}
where $J^2_{\bfn \bfm} = \sum_a \ L^a_{\bfn \bfm} L^a_{\bfn \bfm} = \sum_a \  R^a_{\bfn \bfm} R^a_{\bfn \bfm}$ is the quadratic Casimir operator for the group. Physically this term can be understood as measuring the magnitude of the dimensionless color-electric field on every site. For large dimensionful couplings $g^2_\text{qft}$ this term makes long flux lines energetically unfavorable so that color charges are confined into colorless objects. In this sense the lattice formulation of gauge theories provides a very natural explanation for confinement in gauge theories. Note that the couplings in equations \ref{eq:H_square} and \ref{eq:H_E} are related. This is done so that the continuumlimit respects Lorentz symmetry.
\footnote{One can show that $U^\square$ is related to the QFT field tensor $F$ through ${ U^\square_{\bf n, e_1, e_2} + h.c. = e^{i \ g_\text{qft} a^2 \ F{\bf e_1 e_2}(\bfn) } + h.c. \approx 1 + g^2_\text{qft} a^4 F^2_{\bf e_1 e_2}(\bfn) }$. Since $J$ is related to the dimensionful QFT color-electric field $E$ through $J_{\bfn \bfm} = \frac{a^{D-1}}{g_{qft}} E(\bfn)_{\bfm - \bfn} = \frac{a^{D-1}}{g_{qft}} F(\bfn)_{\text{temporal} \, \bfm - \bfn}$ it is evident that the couplings for $H_\square$ and $H_E$ must be related to regain Lorentz symmetry in the continuum.}

\subsection{Sublattice particle-hole transformation}
\label{section:Sublattice particle-hole transformation}
In what follows our life will be greatly simplified by the particle-hole transformation :
\begin{align}
    \varphi_{\bfn \alpha} =
    \begin{cases}
        \bfn \in \text{even sublattice : } \psi_{\bfn \alpha}  \\
        \bfn \in \text{odd sublattice \hspace{1mm}: } \psi_{\bfn \alpha}^\dagger 
    \end{cases}
\end{align}
so that $H_{\text{hopping}} = H_F + H_F^\dagger$ with 
\begin{align}
    H_F &= \frac{i}{a}\sum_{\substack{ <\bfn \bfm> \\ \bfn \in \text{even} }}  \eta^*_{\bfn \bfm} \varphi_{\bfn \alpha} U^{\dagger \alpha \beta}_{\bfn\, \bfm} \varphi_{\bfm \beta}+ \frac{i}{a}\sum_{\substack{ <\bfn \bfm> \\ \bfn \in \text{odd} }} \eta_{\bfn \bfm} \varphi_{\bfn \alpha} U^{\alpha \beta}_{\bfn \, \bfm} \varphi_{\bfm \beta} \ \ .
\end{align}

With this $H_F$ and $H_F^\dagger$ respectively contain only fermionic annihilation and creation operators so that they act diagonally on fermionic coherent states. Additionally, for further notational convenience we introduce $\Dph_{\alpha\beta}(g) = D^*_{\alpha\beta}(g)$ if $\bfn$ is even and $\Dph_{\alpha\beta}(g) = D_{\alpha\beta}(g)$ if $\bfn$ is odd. With this we define the particle hole transformed gauge field ladder operator :  
\begin{align}
    \Uph^{j; \alpha \beta}_{\bfn \bfm} = \int dg \ \Dph^j_{\alpha \beta}(g) \ket{g} \bra{g} =
    \begin{cases}
        \bfn \in \text{even : }U^{\dagger \ j; \alpha \beta}_{\bfn \bfm} \\
        \bfn \in \text{odd \hspace{1mm}: }U^{j; \alpha \beta}_{\bfn \bfm} 
    \end{cases}\label{eq:ph_transformed_U} \ \ .
\end{align}
Finally, we also define $\alpha_{\bfn \bfm} = \eta^*_{\bfn \bfm}$ if $\bfn$ is even and $\eta_{\bfn \bfm}$ if $\bfn$ is odd. With this we get that $H_{F}$ further simplifies to : 
\begin{align}
    H_{F} = \frac{i}{a}\sum_{<\bfn \bfm>} \alpha_{\bfn \bfm} \varphi_{\bfn \alpha} \Uph^{j;\alpha \beta}_{\bfn \bfm} \varphi_{\bfm\beta} \ \ .
\end{align}

Similarly, the chemical potential and mass terms will transform into : 
\begin{align}
    H_{\text{chemical}} &= \mu \sum_{\bfn} (-1)^{\bfn} \varphi^\dagger_{\bfn \alpha} \varphi_{\bfn \alpha} + \text{some irrelevant constant}
\end{align}
and
\begin{align}
   H_{\text{mass}} &= m \sum_{\bfn} \varphi^\dagger_{\bfn \alpha} \varphi_{\bfn \alpha} + \text{some irrelevant constant} \ \ .
\end{align}

For clarity, we list all the relevant notations below :
\begin{center}
    \begin{tabular}{ l | c | c |c | c }
       & $\varphi^\dagger$ & $\alpha$ & $\Uph$    &  $\Dph$    \\
    \hline \hline
    on even sites/links & $\psi^\dagger$ & $\eta^*$ & $U^\dagger$ &  $D^*$ \\
    on odd sites/links & $\psi$          & $\eta$   & $U$         &  $D  $ \\

\end{tabular}
\end{center}
and note that, from now on, to lighten the notation we will drop the color indices unless they are necessary for clarity.

\section{Foundations of projected entangled pair states (PEPS)}
\label{section:Foundations of projected entangled pair states (PEPS)}
PEPS are a type of tensor network that are particularly well suited to describe states in the Hilbert space of low dimensional quantum many body systems. In this introduction we will focus on the $2+1D$ case, but the generalization to higher dimensions is trivial albeit somewhat denser in notation. To start defining PEPS for quantum many body systems with local physical Hilbert spaces $\mathcal{H}_{\text{physical at } \bfn}$ \emph{(for our intent this will be the matter degrees of freedom)} we must first introduce a new auxiliary Hilbert space $\mathcal{H}_{\text{virtual at }\bfn\text{ towards }\bfei}$ per site $
\bfn$ and direction $\pm \bfei$; this includes the negative unit vectors and since we are working in 2+1D this means we have $4$ of these. Using these auxiliary Hilbert spaces, we define the PEPS on-site state as
\begin{align}
    \ket{T} = \prod_{\bfn} \del{ \sum_{\substack{l \,u \, r \, d \in \text{auxiliary spaces} \\ p \in \text{physical space}  }} T_\bfn^{p \ l \, u \, r \, d} \ket{l u r d}_\bfn \ket{p}_\bfn }
\end{align}
where $l \,u \, r \, d$ respectively denote left, up, right and down. Note the abuse of notation where e.g. $\ket{l}$ indicates the vector labeled by $l$ living in the left virtual Hilbert space. Using the same abuse of notation, we define bond states
\begin{align}
    \ket{B} = \prod_{\bfn} \del{ \sum_{l \,u \, r \, d} \delta_{r\, l} \ket{r}_\bfn \ket{l}_{\bfn + \bfex}  \delta_{u\, d} \ket{u}_
    \bfn \ket{d}_{\bfn+\bfey}  } 
\end{align}
that are simply maximally entangled state between the virtual Hilbert spaces living at either side of each link. Finally, to get the PEPS, which should live only on the physical Hilbert space, we project the entangled pairs onto the on-site state to get
\begin{align}
    \ket{\psi_{PEPS}(T)} = \braket{B|T} 
\end{align}
which reveals why these states are named projected entangled pair states.

In Penrose notation (cfr. appendix \ref{appendix:The Penrose graphical notation})  PEPS are simply : 
\begin{align}
    \ket{\psi_{PEPS}(T)} = \braket{B|T} = \includegraphics[valign = c, height = 2cm]{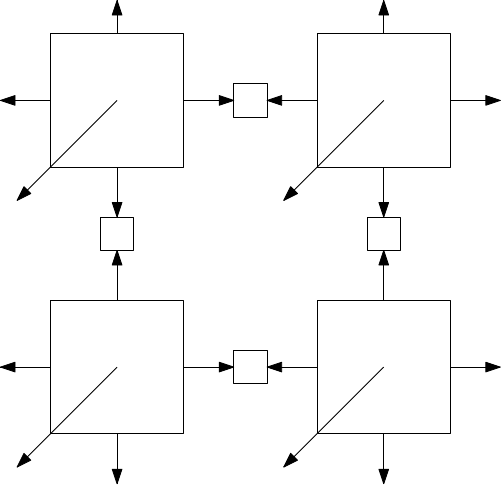}
\end{align}
which is often redefined in terms of the (3,2) tensors $T^{p \ u\, r}_{\bfn \ l \, d} = \delta_{l \tilde{l}}\ \delta_{d \tilde{d}}\ T^{p \ \tilde l \ u \ r \ \tilde d}_{\bfn}$ that are graphically represented as 
\begin{align}
    \includegraphics[valign = c, height = 2.5cm]{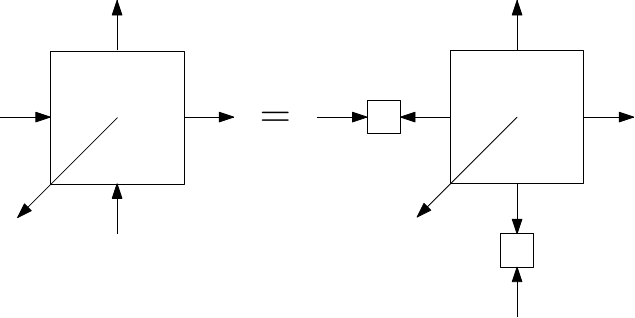} \ \ \ .
\end{align}

As mentioned in the introduction, PEPS with global symmetries can be constructed from local tensors that are representations of that group. To see that this is true, let us denote $U_g$ with $g \in G$ as the local (i.e. on-site) action of the global symmetry. It is then clear that local tensors satisfying
\begin{align}
    U_g T = \includegraphics[valign = c, height = 2cm]{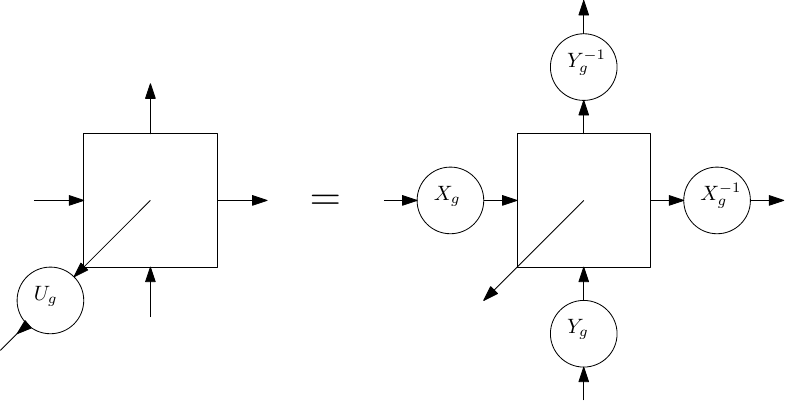}
    \label{eq:PEPS_global_symmetry}
\end{align}
will lead to $\otimes_{\bfn \in lattice} U_{g \ \text{at} \ \bfn }\psi_{PEPS}(T) = \psi_{PEPS}(U_g T) = \psi_{PEPS}(T)$. Furthermore, since $U_g U_h = U_{gh}$ consistency requires $X_g X_h = X_{gh}$ and $Y_g Y_h = Y_{gh}$ so that the X and Y tensors are representations of the symmetry group\footnote{In fact, it is sufficient that $X_g X_h = \omega(g, h) X_{gh}$ which tells us that $X_g$ (and similarly $Y_g$) are also allowed to be projective representations of the group. Indeed, with this we still get that $\psi_{PEPS}(U_g U_h T)$ equals $\psi_{PEPS}(U_{gh} T)$ up to some phase and this is all we really care for.}\cite{Singh_Tensor_network_decompositions_in_the_presence_of_a_global_symmetry}. Consequently, equation \ref{eq:PEPS_global_symmetry} tells us that vectors in the virtual and physical Hilbert spaces of the PEPS tensor carry charges from the global symmetry group. Additionally, the Wigner-Eckart theorem implies that the only nonzero elements in the PEPS tensor are those coupling charges that add up to zero. 

Note that this also implies that the auxiliary Hilbert should be fermionic if one wants to simulate fermionic matter. Luckily it is well known how to deal with this \cite{Kraus_fermionic_projected_entangled_pair_states}.

\subsection{Gauging PEPS}
\label{section:Gauging_PEPS}
In the context of lattice gauge theory, equation \ref{eq:PEPS_global_symmetry} already seems very familiar to the Gauss law. From this perspective it seems that we might want to interpret the group charges in the PEPS auxiliary space as being some sort of counting devices for some purely virtual flux, of Wilson lines, along which the virtual degrees of freedom on the PEPS edges are excited. Equation \ref{eq:PEPS_global_symmetry} then conveys the fact that these, purely virtual, Wilson lines can only terminate on physical charges which indeed ensures that the contracted PEPS carries no net charge so that it is symmetric. 

To gain some intuition, consider a simple example for states with global $Z_2$ symmetry. Here there are two one dimensional irreps $\bf{0}$ and $\bf{1}$, and we will consider states with on-site physical Hilbert space $\mathcal{H}_{phys} = \bf{0} \oplus \bf{1}$. According to the above arguments the on-site PEPS tensor must be a (3,2) tensor $\mathcal{H}_{aux \ \bfex} \otimes \mathcal{H}_{aux \ \bfey} \otimes \mathcal{H}_{phys} \leftarrow \mathcal{H}_{aux \ \bfex} \otimes \mathcal{H}_{aux \ \bfey} $ where $\mathcal{H}_{aux \ \bfex} = \bf{0}^{\otimes n_{0, \bfex}}\oplus\bf{1}^{\otimes n_{1, \bfex}}  $ and similar for the y direction. To keep things simple, let us consider the scenario where $\mathcal{H}_{aux \ \bfex} = \bf{0}$ and $\mathcal{H}_{aux \ \bfey} = \bf{0} \oplus \bf{1}$. With these spaces, one simply example of a PEPS tensor satisfying \ref{eq:PEPS_global_symmetry} is :  
\begin{align}
    T = \includegraphics[valign = c, height = 2cm]{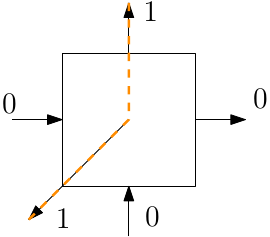} + \includegraphics[valign = c, height = 2cm]{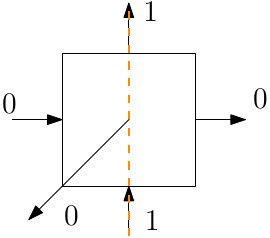} + \includegraphics[valign = c, height = 2cm]{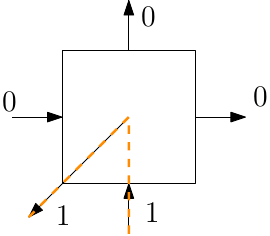}
\end{align}
where we have explicitly added the orange lines to highlight the position of the abovementioned virtual Wilson loops\footnote{With the current auxiliary spaces this is the most general state up to some prefactors for the different contributions. However, when larger auxiliary spaces are considered the possible PEPS become more complicated and physically relevant. Most notable one can write down PEPS that respect the lattice space groups and even conformal symmetry when we allow the amount of auxiliary field to diverge.}. When contracted, the state emerging from this PEPS will be :
\begin{align}
    \ket{\psi_{PEPS}(T)} = \ket{0} + \sum_\bfn \sum_{\Delta_y} \ket{\bfn \ \ \bfn+\Delta_y} + \cdots
\end{align}
where $\ket{0}$ is the $Z_2$ trivial state everywhere and $\ket{\bfn \ \ \bfn+\Delta_y}$ the empty state that is $Z_2$ trivial everywhere except on sites $\bfn$ and $\bfn+\Delta_y$. The dots indicate that there are terms with multiple pairs of particles as well. Most notably, the contracted state knows nothing about these gauge fields but nevertheless they did play a crucial role in the construction of the state.

To make these Wilson loops appear in the final contracted state we use the procedure from \cite{Zohar_Building_projected_entangled_pair_states_with_local_gauge_symmetry}. The first step in this procedure is to associate a Hilbert space with gauge degrees of freedom to every link of the lattice. This is done by supplementing the PEPS on-site state $\ket{T}$ with a reference state that lives in the LGT Hilbert space. In particular, the reference state is chosen to be the electric vacuum $\ket{E=0} = \otimes_{links}\ket{j=0}_{link}$, and with this we get :
\begin{align}
    \ket{T} \in \mathcal{H}_{matter}\otimes\mathcal{H}_{virtual} \ \ \ \rightarrow \ \ \  \ket{T}\ket{E=0} \in \mathcal{H}_{matter} \otimes \mathcal{H}_{gauge} \otimes \mathcal{H}_{virtual} \ \ . 
\end{align} 
Next, we entangle the newly introduced gauge degrees of freedom with the incoming virtual degrees of freedom of each local on-site PEPS tensor (for the 2D case these are the top and right auxiliary Hilbert spaces). To do this, we introduce operators $\hat{U}_{\bfn \, \bfei}$ that act on $\mathcal{H}_{\text{gauge of link emanating from } \bfn \text{ towards } \bfei}  \otimes \mathcal{H}_{\text{virtual of site } \bfn \text{ towards } \bfei} $ and are given by $\hat{U}_{\bfn \, \bfei} = \int dg \ket{g}\bra{g}_{\bfn \bfm} \otimes  e^{i \phi_a(g) \hat{Q}^a_{\bfn \, \bfei}}$. These operators are similar to those defined in \ref{eq:def_U} and an argument similar to the one presented there allows one to show that 
\begin{align}
    [\hat L^a_{\bfn \, \bfei} , \hat{U}_{\bfn \, \bfei} ] = \hat Q^a \, \hat{U}_{\bfn \, \bfei}    
\end{align}
i.e. the $\hat{U}_{\bfn \, \bfei}$ operators entangle the virtual and gauge degrees of freedom. Colloquially speaking, these operators lift the virtual Wilson lines to the physical level. With these, we are finally ready to entangle the gauge degrees of freedom with the PEPS virtual degrees of freedom: 
\begin{align}
    \ket{T}\ket{E=0} \ \ \ \rightarrow \ \ \ \int D\mathcal{G} \ket{\mathcal{G}} U(\mathcal{G}) \bra{\mathcal{G}}  \ \ \  \ket{T}\ket{E=0}
\end{align}
where $D\mathcal{G} = \prod_{links} dg_{link}$, $\ket{\mathcal{G}} = \otimes_{links} \ket{g_{link}}_{link}$ and $U(\mathcal{G}) = \prod_{\bfn \ \bfei} e^{i \phi_a(g_{link}) \hat{Q}^a_{link}}$. Finally, we define the gauged PEPS as the contracted version of this i.e. 
\begin{align}
    \int D\mathcal{G} \ket{\mathcal{G}} U(\mathcal{G}) \bra{\mathcal{G}} \ \ \ \ket{T} \ket{E=0}  \ \ \ \rightarrow \ \ \ \int D\mathcal{G} \ket{\mathcal{G}} \bra{B} U(\mathcal{G}) \ket{T}
\end{align}
where we used the fact that $\braket{G|E=0} = 1$ to simplify the right-hand side. 

The result of this is that the previously hidden gauge degrees of freedom are now explicitly present in the final contracted state. Furthermore, since we can now act on these Wilson lines, the constraint \ref{eq:PEPS_global_symmetry} becomes a true Gauss law generated by \ref{eq:local charge}. The proof of this and many more details on the gauging procedure can be found in \cite{Zohar_Fermionic_projected_entangled_pair_states_and_local_U1_gauge_theories,Zohar_Building_projected_entangled_pair_states_with_local_gauge_symmetry}.

\subsection{Gaussian PEPS and superpositions thereof}
Apart from their theoretical interest as the backbone of the gauging operation described above, PEPS are also often used as a numerical Ansatz for the ground states of quantum many body Hamiltonians. In this context \ref{eq:PEPS_global_symmetry} can be helpful to reduce the number of free parameters in a PEPS with global symmetries. Nevertheless, exactly contracting PEPS remains an NP hard problem and numerical techniques therefore involve approximate contractions of the tensor network. One workaround is to constrain to the subset of PEPS that are built from Gaussian tensors \cite{Mortier_Tensor_Networks_Can_Resolve_Fermi_Surfaces,Shuch_Gaussian_Matrix_Product_States}. In this scenario one can efficiently contract the networks using the covariance matrix techniques introduced by Bravyi \cite{Bravyi_Lagrangian_representation_fermionic_linear_optics}. Since much of our work builds on these techniques, we will also briefly review them in the next section. Note that this is a good moment to look at appendix \ref{appendix:Graphical notation for Gaussian fermionic tensors} as it may help gain some intuition for what these states physically represent.

To start, let us define a Gaussian state in terms of a set of fermionic creation operators $\psi^\dagger_i$ and the empty state $\ket{0}$ as : 
\begin{align}
    \ket{T} = \exp{ \del{ \frac{1}{2} \sum_{ij} T_{ij} \psi^\dagger_i \psi^\dagger_j }} \ket{0}
\end{align}
where $i$ can be any label (e.g. lattice site, color, euclidean time, ...). To work with such states one first defines the Majorana modes 
\begin{align}
    c_{2i} &= \phantom{-i(}\psi_i + \psi^\dagger_i   \\
    c_{2i+1} &= -i(\psi_i - \psi^\dagger_i)
\end{align}
and then defines the covariance matrix 
\begin{align}
(\Gamma_{\rho_T})_{ab} = \frac{i}{2} tr\del{\hat{\rho_T}[\hat c_a, \hat c_b] } = \frac{i}{2} \braket{T \, | [c_a, c_b] \, |T} \ \ \ .
\end{align}
For Gaussian states, Wicks theorem implies that all n-point functions can be obtained from combining 2-point functions (i.e. the entries of the covariance matrix) so that the covariance matrix fully characterizes the state. In \cite{Bravyi_Lagrangian_representation_fermionic_linear_optics,bravyi_gosset_complexity_of_quantum_impurity_problems} it has been shown that operations on these states, or more accurately their density matrices, can be rewritten as essential Linear algebra operations on their covariance matrices.

One such result is that $tr\del{\rho_1 \rho_2} = Pf\del{ \frac{\Gamma_1 + \Gamma_2}{2} }$ which is particularly useful when computing expectation values of PEPS. Indeed, with this we get  
\begin{align}
    \braket{\psi_{PEPS}(T)|\psi_{PEPS}(T)} = \braket{T|B}\braket{B|T} = tr\del{ \rho_{\ket{T}} \del{ \rho_{\ket{B}}\otimes \one }   } = Pf\del{\frac{\Gamma_{\rho_{\ket{T}}} + \Gamma_{\rho_{\ket{B}}\otimes \one }  }{2}} \ \ .
\end{align}
where the $\one$ in $\rho_{\ket{B}} \otimes \one $ is the identity on the physical space. It was introduced to ensure that both density matrices inside the trace act on the same space. The last equality is obviously only true since we are now dealing with PEPS that are built from Gaussian tensors.

Let us now try to set the stage for computation of expectation values of a fermionic operator $\hat{F}$ inside a Gaussian PEPS. Here it is useful to first relate these operators to bit-strings as $\hat{F} = \sum_{a} c^{\text{bit-string}(F)_a} $,  the Hamming weight $w$ of such a string is the number of nonzero elements it contains. For example, when we have three Majorana operators, the bitstring ``110'' represents the operator $c_1 c_2$ and the Hamming weight is $w=2$. With these conventions Bravyi demonstrated that 
\begin{align}
    tr\del{\rho_T \hat{F}} = (-i)^{\text{number of nonzero elements in bitstring}}Pf\del{\rho_T|_{\text{bit-string}}}
\end{align} 
where $\rho_T|_{\text{bit-string}}$ is the $w \times w$ sub-matrix $(\rho_T)_{ab}$ for $a,b$ positions of nonzero elements in the bitstring. Similarly, it was shown that :
\begin{align}
    tr\del{\rho_{T_1} \rho_{T_2} \hat{F}} = tr\del{\rho_{T_1}\rho_{T_1}}Pf\del{i \Delta^\dagger|_{\text{bit-string}} }
\end{align}
where $\Delta = \del{-2 \one + i \Gamma_1 - i \Gamma_2}\del{\Gamma_1 + \Gamma_2}^{-1}$ and this can be used to compute expectation values of products of fermionic operators inside PEPS. Note that all the above results are real and positive definite so that we can replace the Pfaffians with the positive square root of the determinant. Since the latter is numerically cheaper to compute this trick will be applied whenever it is available. Furthermore, these types of computations are explained clearly in more detail in \cite{Shuch_Gaussian_Matrix_Product_States,Zohar_Combining_tensor_networks_with_Monte_Carlo_methods_for_LGT}.

Recently, Bravyi suggested to use states that are superpositions of Gaussians \cite{bravyi_gosset_complexity_of_quantum_impurity_problems} (which may or may not be PEPS). Furthermore, in the same reference it was proven that few Gaussians are needed to capture ground states of Hamiltonians that are quadratic everywhere except in a small localized spatial patch. Intuitively this can be understood as following from the fact that Gaussian states form a basis for all even fermion parity states\footnote{This can demonstrate iteratively by first noting that $\ket{0}$ is obviously Gaussian. Secondly $\psi^\dagger_i \psi^\dagger_j \ket{0} = \exp\del{\psi^\dagger_i \psi^\dagger_j}\ket{0} - \ket{0}$. We can continue like this to show that any fermion even combination can be written as a sum of Gaussian states.} as well as the fact that the ground state of interest will be Gaussian far away from the disturbance. In section \ref{section:Computation of observables} we will apply these new methods to our Ansatz.

\section{The gauged superposition of Gaussian (fermionic) PEPS}
Having discussed the key ingredients, which each in their own right has proven valuable, we are now ready to posit our new Ansatz i.e. gauged PEPS with non-Gaussian bonds described as a superposition of few Gaussian bonds:
\begin{align}
    \ket{\psi}_{GSGFPEPS}  &= \int D\mathcal{G} \ket{\mathcal{G}} \sum_n \lambda_n \bra{B_n} \mathcal{U}(\mathcal{G}) \ket{A} 
\end{align}
where $\ket{\mathcal{G}}$ is a state in the pure gauge Hilbert state in the group representation basis, $\lambda_n$ some complex coefficients, $\bra{B_n}$ a direct product of n-dependent PEPS bond states for each link of the lattice and $\ket{\text{A}}$ a direct product of local PEPS tensors on all sites of the lattice. 

This is generalization of gauged Gaussian PEPS, described in detail in \cite{Kelman_Gauged_Gaussian_projected_entangled_pair_states_A_high_dimensional_tensor_network_formulation_for_lattice_gauge_theories}, which have already proven efficient for a few physical models \cite{Emonts_variational_monte_carlo_simulation_with_tensor_networks_of_a_pure_z3_gauge_theory,Emonts_finding_ground_state_of_lattice_gauge_theory_with_fermionic_tensor_networks,kelman_projected_entangled_pairt_states_for_lattice_gauge_theories_with_dynamical_fermions}. 

As mentioned in the introduction we will first demonstrate how to compute gauge invariant observables for these states and then move on to demonstrate that it is indeed sufficient to only consider $n$-dependent bonds if we are interested in capturing the ground state of lattice gauge theories. Sadly our proof does not imply that few Gaussians are sufficient. However, inspired by the non gauged scenario we expect that our Ansatz is at least able to well approximate LGT states that are close to Gaussian everywhere except on some local spatial patch (and it has been demonstrated that in some cases no superposition is needed and gauged Gaussian PEPS capture the ground state physics well \cite{Emonts_variational_monte_carlo_simulation_with_tensor_networks_of_a_pure_z3_gauge_theory,Emonts_finding_ground_state_of_lattice_gauge_theory_with_fermionic_tensor_networks,kelman_projected_entangled_pairt_states_for_lattice_gauge_theories_with_dynamical_fermions}). Therefore, we suspect the Ansatz to be well suited to simulate isolated bound states such as hadrons and mesons on top of the strongly coupled vacuum. Indeed, for large couplings it is energetically costly to make Wilson lines. Therefore, here the LGT vacuum is $\ket{\text{no particles}}\ket{E=0}$ which is trivially Gaussian. Exited bound states will be more complicated but when localized (by e.g. putting the system inside a box) it should suffice to describe it by superposing few Gaussian states.

\section{Computation of observables}
\label{section:Computation of observables}
In this section we will develop the framework for computing expectation values of gauged superposed Gaussian (fermionic) PEPS. Note that it is straightforward to ignore the gauging operator to get a framework for computing expectation values of superposed Gaussian PEPS. As mentioned before, it is well known that the most efficient way of performing calculations for Gaussian states is through manipulations of the covariance matrices (in Majorana basis) of the corresponding density matrices \cite{Bravyi_Lagrangian_representation_fermionic_linear_optics,bravyi_gosset_complexity_of_quantum_impurity_problems}; therefore the variational parameters for the SGGFPEPS are : $\lambda_n$, $\Gamma_{\rho_{\ket{A}}}$ and $\Gamma_{\rho_{\ket{B_n}}}$.

Before we compute any observables, let us calculate the norm \newline $N = \int D\mathcal{G} \sum_{n m} \lambda^*_n \lambda_m \braket{\psi_n(\mathcal{G})|\psi_m(\mathcal{G})}$ of a SGGFPEPS. To do this, we first focus on : 
\begin{align}
    N_{n m}(\mathcal{G})  = \braket{\psi_n(\mathcal{G})|\psi_m(\mathcal{G})} &= \frac{\braket{A(\mathcal{G})|B_n}\braket{B_n|B_m}\braket{B_m|A(\mathcal{G})} }{\braket{B_n|B_m}} \\
    &= \frac{ tr(\hat\rho_{\ket{A(\mathcal{G})}}\hat\rho_{\ket{B_n}}\hat\rho_{\ket{B_m}})  }{ \braket{ B_n | B_m } }  \\
    &= \frac{ tr(\hat\rho_{\ket{A(\mathcal{G})}} (\one \otimes \hat\rho_{\ket{B_n}}) (\one \otimes\hat\rho_{\ket{B_m}}))  }{ \braket{ B_n | B_m } }  \ \ .
\end{align}

To evaluate the denominator of this expression we use the results in \cite{bravyi_gosset_complexity_of_quantum_impurity_problems} to get : 
\begin{align}
    tr(\hat\rho_{\ket{A(\mathcal{G})}} (\one \otimes \hat\rho_{\ket{B_n}}) (\one \otimes\hat\rho_{\ket{B_m}})) &= (4 i)^{-\frac{n_{ph}+n_{virt}}{2}}pf\del{ 
        \begin{bmatrix}
            i \Gamma_{\rho_{\ket{A(\mathcal{G})}}} & -\one & \one \\
            \one & \begin{bmatrix} 0 & 0 \\ 0 & i \Gamma_{\rho_{B_n}}   \end{bmatrix} & -\one \\
            -\one & \one & \begin{bmatrix} 0 & 0 \\ 0 & i \Gamma_{\rho_{B_m}}   \end{bmatrix}   
        \end{bmatrix}
    } 
\end{align}
where the covariance matrix of the gauged local PEPS tensor is related to the non-gauged one through $\Gamma_{\rho_{\ket{A(\mathcal{G})}}} = O^T(\mathcal{G})\Gamma_{\rho_{\ket{A} }}O(\mathcal{G}) $ where $O(\mathcal{G})$ encodes the action of $U(\mathcal{G})$ on the virtual Majorana fermions i.e. $U(\mathcal{G}) \gamma_a U(\mathcal{G})^\dagger = O(\mathcal{G})_{ab} \gamma_b$. For more details cfr. \cite{Kelman_Gauged_Gaussian_projected_entangled_pair_states_A_high_dimensional_tensor_network_formulation_for_lattice_gauge_theories}.

To evaluate the denominator $\braket{B_n|B_m}$, and in particular its phase,  we need to introduce some reference Gaussian bond state $\ket{B_0}$ with covariance matrix $\Gamma_{\rho_{\ket{B_0}}}$.\footnote{Since the covariance matrices represent density matrices they contain no phase information hence the subtleties in this section !} Since $\lambda_n \rightarrow e^{i \theta_n} \lambda_n$ while simultaneously taking $\Ket{B_n}  \rightarrow e^{i \theta_n} \Ket{B_n}$ leaves the state invariant, we may assume, without loss of generality that $\braket{B_n|B_0} \in \mathbb{R}$ \emph{(this is one of the key insights of \cite{bravyi_gosset_complexity_of_quantum_impurity_problems})}. With this at hand we may write :
\begin{align}
    \braket{B_n|B_m} &= \frac{ \braket{B_0|B_n}\braket{B_n|B_m}\braket{B_m|B_0} }{ \braket{B_0|B_n}\braket{B_m|B_0} } \\
    &= \frac{ (4i)^{-n_{virt}/2} pf\del{ \begin{bmatrix} i\Gamma_{\rho_{B_0}} & \one & \one \\ \one & \Gamma_{\rho_{B_n}} & -\one \\ -\one & -\one & i \Gamma_{\rho_{B_m}} \end{bmatrix} }}{ \sqrt{ pf\del{\frac{\Gamma_{\rho_{B_0}}+\Gamma_{\rho_{B_n}}}{2}} pf\del{\frac{\Gamma_{\rho_{B_0}}+\Gamma_{\rho_{B_m}}}{2}} } } \ \ .
\end{align} 
where we have used $\braket{B_n|B_0} = \sqrt{ tr(\rho_{B_n} \rho_{B_0} ) } = pf\del{ \frac{\Gamma_{\rho_{B_n}} + \Gamma_{\rho_{B_0}}}{2} }$ due to the fact that $\braket{B_n|B_0}$ was chosen to be real.

With this the norm of our state is : $N = \int D\mathcal{G} \  N(\mathcal{G}) =  \int D\mathcal{G} \sum_{n m} \lambda^*_n \lambda_m N_{n m}(\mathcal{G})$ where we have introduced the handy notation $N(\mathcal{G}) = \sum_{n m} \lambda^*_n \lambda_m N_{n m}(\mathcal{G})$ for the norm of the state in a fixed background gauge configuration.

\subsection{Computation of pure Wilson loops}
Using the above, the computation of Wilson loop observables $W$ (i.e. oriented products of $U$ operators along a closed path) is straightforward. Indeed, $U$ is diagonal in the $\ket{\mathcal{G}}$ basis so that $\braket{\mathcal{G}|W|\mathcal{G}'} = \delta(\mathcal{G}, \, \mathcal{G}')W(\mathcal{G})$. With this we get:
\begin{align}
    \braket{W} = \frac{\int D\mathcal{G} \  W(\mathcal{G}) \ N(\mathcal{G})  }{N} = \int D\mathcal{G} \ p(\mathcal{G})\  W(\mathcal{G})
\end{align}
where $p(\mathcal{G}) = N(\mathcal{G}) / N$ is a positive definite probability distribution over gauge field configurations. Therefore, the integral can be computed using Monte Carlo methods without worrying about a sign problem.

\subsection{Fermionic observables}
A generic fermionic LGT observable is of the form $\hat F = \hat W(F)_{ \{\alpha \} } \hat c^{\text{bit-string(F)}}_{ \{\alpha \} } $ in which 
\begin{align}
    c^\text{bit-string(F)}_{ \{\alpha \} } = \sum_{\substack{\bfn \\ \text{type} \in \{1,2\} \\ \alpha \in 1:N_\text{colors} }} c_{\bfn \, \text{type} \, \alpha}^{\text{bit-string}(F)_{\bfn \, type \, \alpha}} = \sum_{i \in 1:2|\text{lattice}|N_\text{color}} c_i^{\text{bit-string(F)}_{i}}
\end{align}
acts on $2|\text{lattice}| N_{\text{colors}}$ Majorana modes $c_{\bfn \, 1 \, \alpha} = \psi_{\bfn \, \alpha}^\dagger + \psi_{\bfn \, \alpha}   $ and $c_{\bfn \, 2 \, \alpha} = i \del{\psi_{\bfn \, \alpha}^\dagger - \psi_{\bfn \, \alpha}}$ and represents a collection of fermionic operators acting on the lattice (\emph{for more details on this notation, we once again refer to \cite{bravyi_gosset_complexity_of_quantum_impurity_problems}}). The second part $W(F)$ represents a configuration of open flux lines connecting these charges to each other and since it is build from $U^{\alpha \, \beta}_{\bfn \, \bfm}$ operators we get that $\braket{\mathcal{G}|W(F)_{\{ \alpha\}}|\mathcal{G}'} = W(F,\mathcal{G})_{\{ \alpha\}} \delta(\mathcal{G} , \,  \mathcal{G}')$. From now the $F$ label will be omitted for clarity. With this notation we get : 
\begin{align}
    \braket{F} &= \frac{\int D\mathcal{G} \  W(\mathcal{G})_{ \{\alpha \} } \  \sum_{n m} \braket{\psi_n(\mathcal{G})| c^\text{bit-string}_{ \{\alpha \} } |\psi_m(\mathcal{G})}  }{N} \\
    &= \frac{\int D\mathcal{G} \  N(\mathcal{G}) \ W(\mathcal{G})_{ \{\alpha \} } \frac{ \sum_{n m} \braket{\psi_n(\mathcal{G})| c^\text{bit-string}_{ \{\alpha \} } |\psi_m(\mathcal{G})} }{N(\mathcal{G})}  }{N} \\
    & = \int D\mathcal{G} \  F(\mathcal{G}) \ p(\mathcal{G}) 
\end{align}
where just like before, $p(\mathcal{G}) = N(\mathcal{G})/N$ and 
\begin{align}
    F(\mathcal{G}) = W(\mathcal{G})_{ \{\alpha \} } \frac{ \sum_{n m} \braket{\psi_n(\mathcal{G})| c^\text{bit-string}_{ \{\alpha \} } |\psi_m(\mathcal{G})} }{N(\mathcal{G})}
\end{align}
is the fermionic observable for a given gauge field configuration. As before, the $\mathcal{G}$ integral can be approximated using sign problem free Monte Carlo integration due to the fact that $p(\mathcal{G})$ is positive definite.

To evaluate $F(\mathcal{G})$ we first rewrite
\begin{align}
    \braket{\psi_n(\mathcal{G})| c^\text{bit-string}_{ \{\alpha \} } |\psi_m(\mathcal{G})} &= \frac{\braket{A(\mathcal{G})|B_n}\bra{B_n}c^\text{bit-string}_{ \{\alpha \} } \ket{B_m}\braket{B_m|A(\mathcal{G})}}{\braket{B_n|B_m}} \ \ . 
\end{align}
and since we already know how to compute $\braket{B_n | B_m}$ let us focus n the numerator. To do this we note that it can be rewritten as : 
\begin{align}
    &tr\del{ (\one \otimes \rho_{B_m}) c^\text{bit-string}_{ \{\alpha \} } \rho_{A(\mathcal{G})} (\one \otimes \rho_{B_n}) } \\ 
    & =(4 i)^{-\frac{n_{ph}+n_{virt}}{2}}pf\del{\begin{bmatrix} \begin{bmatrix} 0 & 0 \\ 0 & i\Gamma_{\rho_{B_n}} \end{bmatrix} & -\one & \one & 0 \\ \one & \begin{bmatrix} 0 & 0 \\ 0 & i\Gamma_{\rho_{B_m}} \end{bmatrix} & -\one & 0 \\ -\one & \one & i D_x \Gamma_{\rho_{A(\mathcal{G})}} D_x^T & -J_x^T + i D_x^T \Gamma^T_{\rho_{A(\mathcal{G})}}J_x^T \\ 0 & 0 & -J_x + i J_x \Gamma_{\rho_{A(\mathcal{G})}} D_x & i J_x \Gamma_{\rho_{A(\mathcal{G})}} J_x^T  \end{bmatrix}} \nonumber
\end{align}
where in the last line we defined $D_x$ and $J_x$ as in \cite{bravyi_gosset_complexity_of_quantum_impurity_problems} and for completeness we repeat their definitions here. To do this we first need to define $w$ as the Hamming weight (i.e. the number of nonzero elements) of the bit-string defining the fermionic operator. With this $J_x$ is the $2N \times w$ matrix with entries $(J_x)_{i,j} = 1$ if j is the position of the i-th nonzero element of the bit-string and zero otherwise. Similarly, $D_x$ is defined as the $2N \times 2N$ diagonal matrix with the i-th diagonal element equal to bit-string$_i$. 

\subsection{Electric field observables}
Computing the electric energy at link $\bfn \bfm$ (with $\bfn$ and $\bfm$ nearest neighbors) involves : 
\begin{align}
    \braket{J^2_{\bfn\,\bfm} } = \frac{\int D\mathcal{G} D\tilde{\mathcal{G}} \braket{\mathcal{G}|J^2_{\bfn\,\bfm}|\tilde{\mathcal{G}}} N(\mathcal{G}) \frac{\sum_{n m} \lambda^*_n \lambda_m \braket{\psi_n(G)}|\psi_m(\tilde{G}) }{N(G)} }{N} \equiv \int DG \, f_E(\mathcal{G}) p(\mathcal{G})
\end{align}
where $p(\mathcal{G})$ is as defined above so that the $\mathcal{G}$ integral can be solved in a sign-problem free fashion. However, 
\begin{align}
    f_E(\mathcal{G}) = \int D\tilde{\mathcal{G}} \braket{\mathcal{G}|J^2_{\bfn\,\bfm}|\tilde{\mathcal{G}}}\frac{\sum_{n m} \lambda^*_n \lambda_m \braket{\psi_n(\mathcal{G})}|\psi_m(\tilde{\mathcal{G}}) }{N(\mathcal{G})} 
\end{align}
which contains a sign-problem full integral over $\tilde{\mathcal{G}}$. Luckily $\braket{\mathcal{G}|J^2_{\bfn \bfm}|\tilde{\mathcal{G}}}$ is only nonzero on the link $\bfn \bfm$. Therefore, if we define $\ket{g_{\bfn \bfm} ; \mathcal{G}} = \ket{\mathcal{G}}$ on all links except the $\bfn \bfm$ link where it is $g_{\bfn \bfm}$ we get : 
\begin{align}
    f_E(\mathcal{G}) = \int dg_{\bfn \bfm} \braket{\mathcal{G}|J^2_{\bfn\,\bfm}|g_{\bfn \bfm} ; \mathcal{G}}\frac{\sum_{n m} \lambda^*_n \lambda_m \braket{\psi_n(\mathcal{G})}|\psi_m(g_{\bfn \bfm} ; \mathcal{G}) }{N(\mathcal{G})} 
\end{align}
in which $\braket{\mathcal{G}|J^2_{\bfn\,\bfm}|g_{\bfn \bfm} ; \mathcal{G}}$ where  is known analytically and $N(\mathcal{G})$ already discussed. All that remains is 
\begin{align}
    \braket{\psi_n(\mathcal{G})|\psi_m(g_{\bfn \bfm} ; \mathcal{G})} = \frac{\braket{A(\mathcal{G})|B_n}\braket{B_n|B_m}\braket{B_m| \mathcal{U}(g_{\bfn \bfm}) |A(\mathcal{G})} }{\braket{B_n|B_m}} 
\end{align} 
which can be computed since $\braket{B_n|B_m}$ is already known and $\mathcal{U}(g_{\bfn \bf \bfm})$ is just some operator acting on the virtual degrees of freedom so that we may write $\mathcal{U}(g_{\bfn \bfm})=c^{\text{different bit-string}(g_{\bfn \bfm})}$. This allows us to again use the techniques from \cite{bravyi_gosset_complexity_of_quantum_impurity_problems} to get :
\begin{align}
    \braket{A(\mathcal{G})|B_n}\braket{B_n|B_m}&\braket{B_m| \mathcal{U}(g_{\bfn \bfm}) |A(\mathcal{G})} \nonumber \\ 
    &= tr\del{ \rho_{A(\mathcal{G})} (\one \otimes \rho_{B_n}) (\one \otimes \rho_{B_n}) c^{\text{different bit-string}(g_{\bfn \bfm})} }
\end{align}
which can be computed efficiently using the above described techniques.

\section{Exact representation of the LGT ground state as a gauged superposition of Gaussian PEPS}
In this section we will take a first step towards demonstrating the expressive power of the Ansatz by showing that the ground state of any LGT can be written as a gauged superposition of Gaussian PEPS. As mentioned before, we are unable to prove that the amount of states in the superposition is small but nevertheless this section is required as it ensures that it is sufficient to only consider a superposition of PEPS with distinct bonds. 

To achieve this goal, we use the fact that the ground state of any gapped\footnote{If the Hamiltonian is not gapped we can always artificially add a small mass term which we later take to zero. Additionally, since we are eventually interested in simulating the standard model, which contains massive fermions, this poses no real problem. } Hamiltonian can be obtained through imaginary time evolution \cite{Peskin_Introduction_to_QFT} i.e.:
\begin{align}
    \ket{\psi_0} = \lim_{\beta \rightarrow \infty} e^{-\beta H} \ket{\psi_\text{intial}} = \lim_{\substack{\beta \rightarrow \infty \\ \epsilon \rightarrow 0}}\underbrace{ e^{-\epsilon H} e^{-\epsilon H}  \cdots e^{-\epsilon H}  }_{\text{$N = \beta/\epsilon$ times }}  \ket{\psi_\text{intial}}  \ \ .
\end{align} 
To see that this can be recast as a gauged Gaussian fermionic PEPS we introduce gauge and fermionic resolutions of the identity around each $e^{- \epsilon H}$ 
\begin{align}
    \ket{\psi_0} &= \lim_{\substack{\beta \rightarrow \infty \\ \epsilon \rightarrow 0}} \ \overbrace{\one_{t = 0} e^{-\epsilon \hat H} \one_{t=1} e^{-\epsilon \hat H} \one_{t=2} \cdots \one_{t = N-1} e^{-\epsilon \hat H} \one_{t=N} }^{\text{$N = \beta/\epsilon$ times }}  \ket{\psi_\text{intial}} \ \ .
\end{align}

Next, we choose to express the fermionic and bosonic parts of these identities using fermionic coherent states and gauge group elements respectively (an introduction to fermionic coherent states and the related topic of Grassmann variables can be found in appendix \ref{appendix:Fermionic coherent states and Grassmann variables}). To achieve this, we introduce a set of Grassmann variables $\varphi_{\bfn \, t}$, thereto related fermionic coherent states $\ket{\varphi_{\bfn \, t}}$ and gauge group elements $\ket{g_{\bfn \, t}}$ for each site $\bfn$ and time $t$. With this we can rewrite the resolution of the identity as : 
\begin{align}
    \one_t = \int D\mathcal{G}_t \ket{\mathcal{G}_t}\bra{\mathcal{G}_t} \otimes \int D\varphi_t D\bar\varphi_t \ket{\varphi_t}\bra{\varphi_t}  e^{-\sum_\bfn \bar\varphi_{\bfn \, t} \ \varphi_{\bfn \,  t} }
\end{align}
where $\ket{\mathcal{G}_t}$ and $\ket{\varphi_t}$ respectively represent gauge and fermionic field configurations on the entire spatial lattice at time $t$ and $\int D\mathcal{G}_t$ and $\int D\bar\varphi_t D\varphi_t$ represent integration over all possible gauge and fermionic (i.e. Grassmann) configurations at that time. Similarly, in the next equation we introduce $\int [D\mathcal{G}]$ and $\int [D\bar\varphi D\varphi]$ that represent integration over gauge and fermionic configurations at all times and places. Using this definition of the identity and dropping the $\lim_{\substack{\beta \rightarrow \infty \\ \epsilon \rightarrow 0}}$ to lighten the notation leads to :  
\begin{align}
   \ket{\psi_0} &= \int [D\mathcal{G}] [D\bar\varphi] [D\varphi] e^{- \sum_{\bfn} \bar\varphi_{\bfn\, t=0}\ \varphi_{\bfn\, t=0}} \ket{\mathcal{G}_0}\ket{\varphi_0}  \label{eq:first_thing} \\
   &\hspace{2cm} \times 
   \del{\prod_{t = 0}^{N-1} e^{- \sum_{\bfn} \bar\varphi_{\bfn\, t}\ \varphi_{\bfn\, t}} \bra{\mathcal{G}_t \varphi_t} e^{-\epsilon H} \ket{\mathcal{G}_{t+1} \varphi_{t+1}}} \braket{\mathcal{G}_N \Theta_N| \psi_\text{initial}} \nonumber \\
   &= \int [D\mathcal{G}] \ket{\mathcal{G}_0} \del{\prod_{t=0}^{N-1} \braket{\mathcal{G}_t | e^{-\epsilon \del{  H_E +  H_\square + H_\square^\dagger }} | \mathcal{G}_{t+1}}} \\
    &\hspace{2cm} \times \int [D\varphi][D\bar \varphi ] e^{-\sum_{\bfn \, t} \bar \varphi_{\bfn \,t} \ \varphi_{\bfn \, t} } \ket{\varphi_0} \del{\prod_{t=0}^{N-1} \braket{\varphi_t | e^{-\epsilon \del{  H_\text{hopping}(\mathcal{G}_t) + H_m +  H_\mu }} | \varphi_{t+1}}  \nonumber} \\
    &= \int [D\mathcal{G}] \ket{\mathcal{G}_0} \ \del{\prod_{t=0}^{N-1} \braket{\mathcal{G}_t| e^{-\epsilon  H_E} |\mathcal{G}_{t+1} }} \  \psi_I([\mathcal{G}]) \otimes \ket{\psi([G])}_{II}
\end{align}
which contains functions $\psi_I([\mathcal{G}]) : [\mathcal{G}] \rightarrow \mathbb{R}$ and $\ket{\psi([\mathcal{G}])}_{II} : [\mathcal{G}] \rightarrow \emph{fermionic Fock space} $ that we will define in sections \ref{sec:pure_gauge} and \ref{sec:fermions} respectively. The analytical derivation is somewhat cumbersome and is therefore relegated to appendix \ref{appendix:technical details of the pure gauge PEPS construction}. Note that, unlike the gauged PEPS introduced in the introduction, the current gauging operators depend on the physical gauge field configuration $\mathcal{G}_0$ as well as a set of virtual (past) gauge field configurations. In section \ref{sec:virtual_gauge_fields_to_non_Gaussianity} we will show that these past gauge field configurations can be eliminated at the cost of introducing non-Gaussianity in the bonds. 
\subsection{The pure gauge part}
\label{sec:pure_gauge}
To write the pure Gauge part of the LGT ground state 
\begin{align}
    \psi_I([\mathcal{G}]) = \prod_t \braket{\mathcal{G}_t | e^{-\epsilon \del{ H_\square + H_\square^\dagger }} | \mathcal{G}_{t+1}}
\end{align}
as a gauged PEPS we introduce two virtual modes $\varphi^{n_\square \, \curvearrowleft}_{\bfn \, \bfei \bfej \, t} $ and $\varphi^{n_\square \, \curvearrowright}_{\bfn \, \bfei \bfej \, t} $  per site $\bfn$, direction pair $\bfei \bfej$ and virtual time $t$. The final label, $n_\square \in 1:n_\square^{max}$ is related to the accuracy of the approximation and in order to get good results we need $n^{max}_\square \gg \frac{\epsilon}{a}\frac{a^{D-3}}{g^2_\text{qft}}$. Using these modes we define the local PEPS tensor as :
\begin{align}
    \ket{A}_I =& \prod_{\substack{\bfn \ t \\ n_\square \in 1:n_\square^{max}} } \exp\del{ \del{\frac{\epsilon}{a n_\square}\frac{a^{D-3}}{g^2_\text{qft}}}^{1/4} \sum_{\substack{\bfei \\ \bfej != \pm \bfei  }}  \varphi^{n_\square \, \curvearrowright \dagger}_{\bfn \, \bfei \, \bfej \, t}\varphi^{n_\square \, \curvearrowright \dagger}_{\bfn \, \bfej \, \bfei \, t} + \varphi^{n_\square \, \curvearrowleft \dagger}_{\bfn \, \bfei \, \bfej \, t}\varphi^{n_\square \, \curvearrowleft \dagger}_{\bfn \, \bfej \, \bfei \, t} }\ket{0}_I 
\end{align}
and for a single time, site and $n_\square$ (whom indices we will suppress) this is visualized as 
\begin{align}
    \includegraphics[width = 7cm]{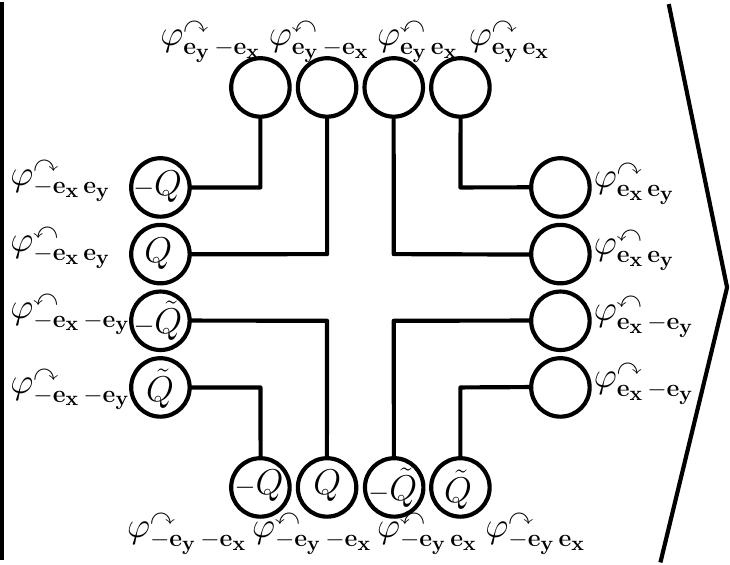}  \label{eq:graphical_fidicual_string_net}
\end{align}
using the graphical notation of appendix \ref{appendix:Graphical notation for Gaussian fermionic tensors}. Similarly to the PEPS described in \ref{section:Foundations of projected entangled pair states (PEPS)}, the bond state is simply a direct product of maximally entangled pairs on each link. In terms of the current notation it is given by
\begin{align}
    \bra{B}_I &= \bra{0}_I \prod_{\braket{\bfn \bfm} \ t} \exp\del{ \sum_{\substack{\bfej != \pm (\bfm-\bfn) \\ n_\square \in 1:n_\square^{max} }}  \varphi^{n_\square \, \curvearrowright}_{\bfn \, \bfm-\bfn \,\bfej}\varphi^{n_\square \, \curvearrowright}_{\bfm \, \bfn-\bfm \,\bfej} + \varphi^{n_\square \, \curvearrowleft}_{\bfn \, \bfm-\bfn \, \bfej}\varphi^{n_\square \, \curvearrowleft}_{\bfm \, \bfn-\bfm \, \bfej}  } \ \ .
\end{align}
Finally, we gauge this PEPS with 
\begin{align}
    \mathcal{U}([\mathcal{G}])_I = \prod_t \mathcal{U}(\mathcal{G}_t)_I = \prod_{\bfn \, t} \exp\del{ i\phi_a(g_{\bfn-\bfex \, \bfn}(t)) Q^{a \, left}_{\bfn \, t} } \exp\del{ i\phi_a(g_{\bfn-\bfey \, \bfn}(t)) Q^{a \, bottom}_{\bfn \, t} }
\end{align}
where the charges are defined by 
\begin{align}
    Q^{a \, left}_{\bfn \, t} &= \sum_{n_\square} \del{ -Q^{a \, \curvearrowright \, n_\square}_{\bfn \, -\bfex \, \bfey \, t} + Q^{a \, \curvearrowleft \, n_\square}_{\bfn \, -\bfex \, \bfey \, t} + \tilde Q^{a \, \curvearrowright \, n_\square}_{\bfn \, -\bfex \, -\bfey \, t} - \tilde Q^{a \, \curvearrowleft \, n_\square}_{\bfn \, -\bfex \, -\bfey \, t} }
\end{align}
and
\begin{align}
    Q^{a \, bottom}_{\bfn \, t} &= \sum_{n_\square} \del{ -Q^{a \, \curvearrowright \, n_\square}_{\bfn \, -\bfey \, -\bfex \, t} + Q^{a \, \curvearrowleft \, n_\square}_{\bfn \, -\bfey \, -\bfex \, t} + \tilde Q^{a \, \curvearrowright \, n_\square}_{\bfn \, -\bfey \, \bfex \, t} - \tilde Q^{a \, \curvearrowleft \, n_\square}_{\bfn \, -\bfey \, \bfex \, t} }
\end{align}
with $Q^a$ and $\tilde Q^a$ as defined in appendix \ref{appendix:Review on relevant group actions}. For further clarity, the type of generators (i.e. $Q$ or $\tilde Q$) and sign in front of it are also indicated in the graphical representation \ref{eq:graphical_fidicual_string_net}. In appendix \ref{appendix:technical details of the pure gauge PEPS construction} we demonstrate that the above is true. 

Although this might look daunting, it is in fact a very logical result. For a single time and $n_\square$ the contracted PEPS will be a superposition of all closed virtual fermion loops that run around a single plaquette. The gauging operator then turns these into physical Wilson lines and the chosen charges do this in a way that leads to a clockwise and counterclockwise oriented Wilson loop. Finally, the effect of the many times and $n_\square$ is to superpose smaller Wilson loops into larger ones.

\subsection{The fermionic part}
\label{sec:fermions}
Similarly, to write the fermionic part of the LGT ground state
\begin{align}
    \ket{\psi([\mathcal{G}])}_{II} &= \int [D\varphi][D\bar \varphi ] e^{-\sum_{\bfn, t} \bar \varphi_{\bfn ,t} \varphi_{\bfn, t} } \ket{\varphi_0} \prod_t \braket{\varphi_t | e^{-\epsilon \del{  H_\text{hopping}(\mathcal{G}_t) + H_m + H_\mu }} | \varphi_{t+1}} \label{eq:first_def_fermionic_wavefunction}
\end{align}
as a Gauged Gaussian PEPS $ \ket{\psi([\mathcal{G}])}_{II} = \braket{B| \mathcal{U}([\mathcal{G}]) |A}_{II}$ we introduce two virtual modes $\varphi^{C\dagger}_{\bfn \, \bfei \, t}$ and $\varphi^{D\dagger}_{\bfn \, \bfej \, t}$ per site, direction and imaginary time. With those we define the on-site PEPS state :
\begin{align}
    \ket{A}_{II} = \prod_\bfn \exp\del{ \varphi^\dagger_{\bfn \, t=0}\sum_{\substack{t=0 \\ \bfei}}^{N-1}\del{r_\bfn \sqrt{\frac{\epsilon}{a}} }^t \varphi^{C\dagger}_{\bfn \, \bfei \, t} + \sum_{\substack{0 \leqslant t \leqslant s \leqslant N-1\\ \bfei \bfej}} \del{ r_\bfn \frac{\epsilon}{a} }^{t-s} \varphi^{C\dagger}_{\bfn \, \bfei \, t} \varphi^{D\dagger}_{\bfn \, \bfej \, s} } \ket{0 }\ket{0}_{II}
\end{align}
where $r_\bfn = 1 + \epsilon m + (-1)^\bfn \epsilon \mu$. Additionally, we define the bond states
\begin{align}
    \bra{B}_{II} = \bra{0}_{II}\prod_{t \braket{\bfn \bfm}} \exp\del{-i \alpha^*_{\bfn \bfm} \varphi^C_{\bfn \, \bfm-\bfn \, t}\varphi^C_{\bfm \, \bfm-\bfn \, t } -i \alpha_{\bfn \bfm} \hat \varphi^D_{\bfn \, \bfm-\bfn \, t}\hat \varphi^D_{\bfm \, \bfm-\bfn \,t} }
\end{align}
that are again just maximal entangled pairs on each link, one subtlety is that the hopping phase $\alpha^*_{\bfn \bfm}$ defined in the end of section \ref{section:Sublattice particle-hole transformation} now appears in the bond states. Finally, the gauging operation is  
\begin{align}
    \mathcal{U}([\mathcal{G}])_{II} = \prod_t \mathcal{U}(\mathcal{G}_t)_{II}  = \prod_{\braket{\bfn \bfm}\, t} e^{i\phi_a(g_{\bfn \bfm}(t)) \del{Q^C_{\bfn \bfm \,t} + Q^D_{\bfn \bfm \,t} }  }
\end{align}
where \begin{align}
    Q^C_{\bfn \bfm \,t} &= 
    \begin{cases}
        -\varphi^{\dagger \, C}_{\bfm \, \bfn-\bfm \,t} T^a \varphi^{ \, C}_{\bfm \, \bfn-\bfm \,t} \text{ on even links} \\
        \varphi^{\dagger \, C}_{\bfm \, \bfn-\bfm \,t} (T^a)^T \varphi^{ \, C}_{\bfm \, \bfn-\bfm \,t} \text{ on odd links}
    \end{cases} \\
    Q^D_{\bfn \bfm \,t} &= 
    \begin{cases}
        \varphi^{\dagger \, D}_{\bfm \, \bfn-\bfm \,t} (T^a)^T \varphi^{ \, D}_{\bfm \, \bfn-\bfm \,t} \text{ on even links} \\
        -\varphi^{\dagger \, D}_{\bfm \, \bfn-\bfm \,t} T^a \varphi^{ \, D}_{\bfm \, \bfn-\bfm \,t} \text{ on odd links}
    \end{cases} \ \ .
\end{align}
in appendix \ref{appendix:technical details of the fermionic PEPS construction} we show that the above is true.

To Intuitively understand this PEPS we realize that the physical matter fields only couple to the type-C virtual fields and that the type-C virtual fields only couple to the type-D fields. With this in mind we can convince ourselves that the contracted PEPS will be a superposition of open Wilson lines that start/end on even/odd sites respectively. Since we are working with staggered fermions this means that the state is a superposition of Wilson lines between + and - particles, as is to be expected. 

\subsection{Eliminating past gauge field configurations}
\label{sec:virtual_gauge_fields_to_non_Gaussianity}
At this point we have proven that the ground state of any LGT can be written as :
\begin{align}
    \ket{\psi_0} = \int [D\mathcal{G}] \ket{\mathcal{G}_0} \ \prod_t \braket{\mathcal{G}_t| e^{-\epsilon  H_E} |\mathcal{G}_{t+1} } \  \psi_I([\mathcal{G}]) \otimes \ket{\psi([\mathcal{G}])}_{II} \label{eq:generalized_gauged_gaussian_fermionic_peps}
\end{align}
with $\psi_I([\mathcal{G}])$ and $\ket{\psi([\mathcal{G}])}_{II}$ as defined above. It is worth noting that one could use Monte Carlo techniques to directly calculate expectation values from this expression. In fact such an approach would be very similar to the conventional path integral based approach to LGT. The only difference is that the current approach breaks up the fermionic part into a Gaussian Tensor network. Obviously, this approach is powerful, but sadly just like conventional Monte Carlo such a method would still suffer from a sign problem. 

To work around this we now sacrifice the Gaussianity in favor of eliminating the past gauge field configurations and therefore the sign problem. More precisely, we will write the above LGT ground state as 
\begin{align}
    \ket{\psi_0}  &= \int D\mathcal{G} \ket{\mathcal{G}} \sum_n \lambda_n \bra{B_n} \mathcal{U}(\mathcal{G}) \ket{A}  \label{eq:desired_form}
\end{align}
which can be though of as resummation of the integral over past gauge field configurations. To get started, define 
\begin{align}
    \mathcal{U}_{t} = \int D\mathcal{G} \ket{\mathcal{G}_t}\bra{\mathcal{G}_t} \otimes \mathcal{U}(\mathcal{G}_t)_{I} \otimes \mathcal{U}(\mathcal{G}_t)_{II} \ \ .
\end{align} 
so that we may write 
\begin{align}
    \ket{\psi_0}  &= \int D\mathcal{G} \ket{\mathcal{G}}\bra{\mathcal{G}} \bra{B}\del{ \prod_{t=0 
    \cdots N-1} e^{- \epsilon H_E }  \mathcal{U}_{t} } \ket{E=0} \ket{A} \label{eq:gen_to_sup}
\end{align}
with $\ket{E=0}$ the vacuum state for the color electric field, $\ket{A} = \ket{A}_{I}\otimes\ket{A}_{II}$ and  $\bra{B} = \bra{B}_{I}\otimes\bra{B}_{II}$.

Additionally, we define
\begin{align}
    Q^a_{\text{all}\, \bfn \bfm \, t} &=+ Q^{a \, \bfn - \bfm}_{\bfm \,t} + Q^C_{\bfn \bfm \, t} + Q^D_{\bfn \bfm \, t}  
\end{align}
which measures the total charge of all types of virtual fermions involved in link $\bfn \bfm$ at time $t$. With this, the gauging operator for that link and time can be written as : 
\begin{align}
    \mathcal{U}_{\bfn \bfm \, t} &= \int dg_{\bfn \bfm}(t) \ket{g_{\bfn \bfm}(t)}\bra{g_{\bfn \bfm}(t)}   e^{i \phi_a(g_{\bfn \bfm}(t)) Q^a_{\text{all}\, \bfn \bfm \, t} }
\end{align}
and it is easy to see that : 
\begin{align}
    \mathcal{U}_{\bfn \bfm \, t}^\dagger L^a_{\bfn \bfm} \mathcal{U}_{\bfn \bfm \, t} = L^a_{\bfn \bfm} + Q^a_{\text{all}\, \bfn \bfm \, t}
\end{align}
which is reminiscent of the commutation relations of the ladder operators below equation \ref{eq:U_transform}.

With this we can commute all $e^{-\epsilon H_E}$ in \ref{eq:gen_to_sup} to the right where they will act on the vacuum and vanish. This leads to : 
\begin{align}
    \ket{\psi_0}  &= \int D\mathcal{G} \ket{\mathcal{G}}\bra{\mathcal{G}} \bra{B} \prod_{\bfn \bfm}\del{\prod_{t=1}^{N-1} e^{ -\frac{\epsilon}{a}\frac{2g^2_\text{qft}}{a^{D-3}} \del{  Q^a_{\text{all before }t \, \bfn \bfm }  }^2 }} \del{\prod_{t=1}^{N-1} \mathcal{U}_t}  \ket{A} \ket{E=0}  
\end{align}
where $Q^a_{\text{all before } t \, \bfn \bfm} = \sum_{s=t}^{N-1} Q^a_{\text{all}\, \bfn \bfm \, s}$. As expected, the above state is no longer Gaussian due to the $\del{Q^a_{\text{all before } t \, \bfn \bfm}}^2$. However, since $\del{Q^a_{\text{all before } t \, \bfn \bfm}}^2$ acts purely on the bonds we can rewrite the resulting non-Gaussian bond state as a superposition over Gaussian bond states which immediately proves \ref{eq:desired_form} and the statement that the ground state of any LGT can be written as a gauged superposition of Gaussian PEPS. 

In principle the exact $\lambda_n$ and $\bra{B_n}$ can be computed exactly but since we are only interested in using the above states as a variational Ansatz this gives no additional advantage. Indeed, to work with these states one should truncate the amount of virtual modes and PEPS in the superposition and then use the results from \ref{section:Computation of observables} together with gradient descent to optimize the variational parameters. Note that this has already been done in the scenario where a single Gaussian PEPS is gauged cfr. \cite{Emonts_variational_monte_carlo_simulation_with_tensor_networks_of_a_pure_z3_gauge_theory}. 

\section{Conclusion and outlook}
We developed a framework for calculation of gauge invariant observables in gauged superposed Gaussian PEPS in the scenario that all PEPS share the same on-site tensors. We then proved that this assumption is sufficient to capture any LGT ground state. 

In the case where few Gaussian states sit in the superposition, computation of observables can be done efficiently. Therefore, in future works we want to better understand when this is the case. In particular the results in \cite{bravyi_gosset_complexity_of_quantum_impurity_problems} seem to indicate that this might be the case for states that are non Gaussian only for a small region in spacetime. One particularly interesting example of this are bound states such as the hadrons and mesons in QCD. 

In the same spirit it would be interesting to obtain bounds on the convergence of the ground state as a function of the various parameters in the Ansatz ($\epsilon, N = \beta/\epsilon$ and $n_\square$) and Hamiltonian ($g_\text{qft}, m$ and $\mu$). 

In the future we also aim to extend this work by considering a continuumlimit of the construction. The resulting Ansatz will be some gauged continuous tensor network and might be useful for the study of continuum gauge theories. 

Finally, we note that one might also consider the non-gauged version of our Ansatz states as a way to approximate the ground state of a fermionic lattice models without gauge fields. Such states were already proposed as a sidenote in \cite{Boutin_quantum_impurity_models_using_superpositions_of_fermionic_Gaussian_states} and are expected to work exceptionally well for quantum impurity problems. 

\acknowledgments
We acknowledge valuable discussions with Umberto Borla, Ariel Kelman, Patrick Emonts and Bastiaan Aelbrecht. This research is funded by the European Union (ERC, OverSign, 101122583). Views and opinions expressed are however those of the author(s) only and do not necessarily reflect those of the European Union or the European Research Council. Neither the European Union nor the granting authority can be held responsible for them.

\appendix
\section{Implementing group actions in second quantization}
\label{appendix:Review on relevant group actions}
Given a group with generators $(T^a)^j$ that satisfy the group algebra $[T^a, T^b] = i f^{abc} T^c$, we can create a representation on the fermionic operators $\psi^\dagger_\alpha$ by defining the charges $Q^a = \psi^\dagger_\alpha (T^a)^j_{\alpha \beta} \psi_\beta$ and group elements $\theta(g) = e^{i \phi_a(g) Q^a}$. Indeed, with these definitions we find : 
\begin{align}
    \theta(g) \psi^\dagger_\alpha \theta^\dagger(g) &= \psi^\dagger_\alpha + i \phi_a(g) \ [Q^a, \psi^\dagger_\alpha] \\
    &= \psi^\dagger_\alpha + i \phi_a(g) \ \psi^\dagger_\beta (T^a)^j_{\beta \alpha} \\
    &= \psi^\dagger_\beta \ \exp\del{ i \phi_a(g) (T^a)^j  }_{\beta \alpha} \\
    &= \psi^\dagger_\beta \ D^j_{\beta \alpha}(g) \ \label{right regular} .
\end{align}
so that $Q^a = \psi^\dagger_\alpha (T^a)^j_{\alpha \beta} \psi_\beta$ generates a right regular linear map on the creation operators $\psi^\dagger_\alpha$. Similarly, 
\begin{align}
    \theta(g) \psi_\alpha \theta^\dagger(g) = \bar D^j_{\beta \alpha}(g)_{\alpha \beta} \psi_\beta = D^j_{\beta \alpha}(g^{-1})_{\alpha \beta} \psi_\beta 
\end{align}
implements a left inverse group action on the annihilation operators. 

Alternatively we might define charges $\tilde{Q}^a = \psi^\dagger_\alpha (T^a)^j_{\beta \alpha} \psi_\beta$ that generate the group $\tilde \theta(g) = e^{i \phi_a(g) \tilde{Q}^a}$, it is easy to show that
\begin{align}
    \tilde \theta(g) \psi^\dagger_\alpha \tilde \theta^\dagger(g) &= D^j_{\alpha \beta}(g) \psi^\dagger_\beta \ \ \ 
\end{align} 
so that $\tilde{\theta}(g)$ acts as a left regular linear map on the creation operators. Similarly,
\begin{align}
    \tilde \theta(g) \psi_\alpha \tilde \theta^\dagger(g) &= \psi_\alpha D^j_{\beta \alpha }(g^{-1}) \ \ \ 
\end{align}
implements a right inverse group action on the annihilation operators.

It is important to note is that the $Q^a$ generators satisfy the group algebra ${[Q^a, Q^b] = i f^{abc} Q^c}$ whereas the $\tilde{Q}^a$ generators satisfy the opposite algebra ${[\tilde{Q}^a, \tilde{Q}^b] = - i f^{abc} \tilde{Q}^c}$. In fact this property can be used to determine if the action of the exponentiated generator will act on the left or on the right of the creation operators $\psi^\dagger$.

\section{The Penrose graphical notation}
\label{appendix:The Penrose graphical notation}
For completeness, we review the relevant part of Penrose his graphical notation for tensors. In this notation a (p, q) tensor $\sum_{\{i, j\}}T^{i_1 \cdots i_p}_{j_1 \cdots j_q} \ket{i_1 \cdots i_p}\bra{j_1 \cdots j_q}$ is represented by a node with $p$ outgoing and $q$ incoming lines. For example a (3,2) tensor could be represented as : 
\begin{align}
    T^{ijk}_{lm} \ \ket{i \, j \, k}\bra{l \, m}= \includegraphics[valign=c, height = 1cm]{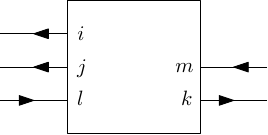} 
\end{align}
and in the same language a contraction is represented by simply connecting the corresponding lines. For example :
\begin{align}
    U_i^a\, T^{ijk}_{mn} \  \ket{a \ j \ k} \bra{l \ m} = \includegraphics[valign=c, height = 1cm]{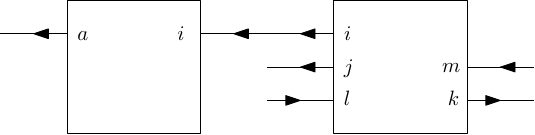} 
\end{align}

For our purposes it is also very convenient to consider the maximally entangled tensors $\ket{B} = \sum_{ij} B_{ij} \ket{i \, j} = \sum_i \ket{i \, i}$ and $\bra{B} = \sum_{ij} B^{ij} \bra{i j}$ (i.e. $B_{ij} = B^{ij} = \delta_{ij}$). These tensors act like a metric in the sense that they allow us raise or lower indices on a tensor, for example $T^{jk}_{alm} = \sum_a B_{ai}\ T^{ijk}_{lm}$ becomes : 
\begin{align}
\includegraphics[valign=c, height = 1cm]{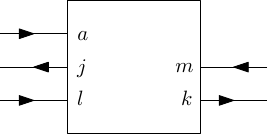} = \includegraphics[valign=c, height = 1cm]{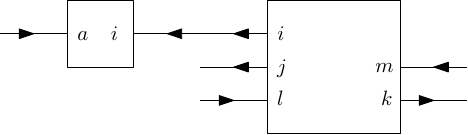} 
\end{align}
Furthermore, since $B_{ij}B^{jk} = \delta_i^k$ we can re-interpret contractions of tensors as projections of the relevant legs onto the metric (i.e. maximally entangled states). For example $\sum_i \, U_i^a\, T^{ijk}_{mn} = \sum_{i\bar{i}} \, B^{i \bar{i}}U_i^a\, T^{jk}_{\bar{i}mn}$. This will be a crucial trick when gauging PEPS.

\section{Graphical notation for Gaussian fermionic tensors}
\label{appendix:Graphical notation for Gaussian fermionic tensors}
In what follows we will introduce a graphical notation which represents Gaussian states $  { \ket{T} = \exp\del{ \frac{1}{2} \sum_{ij } T_{ij} \  \psi^\dagger_i \psi^\dagger_j } \ket{0} } $ and contractions thereof as operations on an associated directed graph with adjacency matrix $T_{ij}$.    

To see how this works, let us first rewrite $\ket{T}$ as : 
\begin{align}
    \ket{T} &= \del{ 1 + \sum_{i > j} T_{ij} \  \psi^\dagger_i \psi^\dagger_j + \sum_{\substack{ i > j\\ k>l }} T_{ij} T_{kl} \  \psi^\dagger_i \psi^\dagger_j \psi^\dagger_k \psi^\dagger_l + \cdots } \ket{0} \label{eq:Gaussian_state}
\end{align}
and focus on the concrete example with $i \in 1,2,3,4$. In this case, the graphical notation for $\ket{T}$ is : 
\begin{align}
    \ket{T} = \includegraphics[valign=c, width = 4cm]{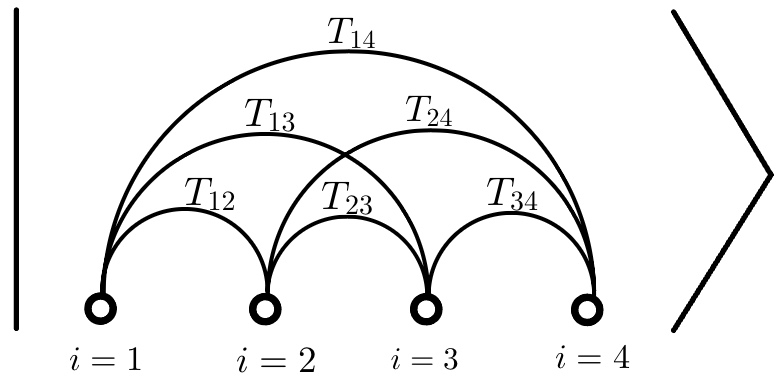}
\end{align} 
where $\ket{\text{adjacency matrix}}$ is to be interpreted as the sum of colored graphs (i.e. non-entangled states) that can be reached by coloring (i.e. occupying) pairs of nodes (i.e. fermionic modes) that are connected by an edge. The prefactor for one coloring choice simply corresponds to product of the weights of the colored lines. Indeed, with these rules we get that $\ket{T}$ is : 
\begin{align}
    \ket{T} =& \underbrace{\includegraphics[valign=c, width = 4cm]{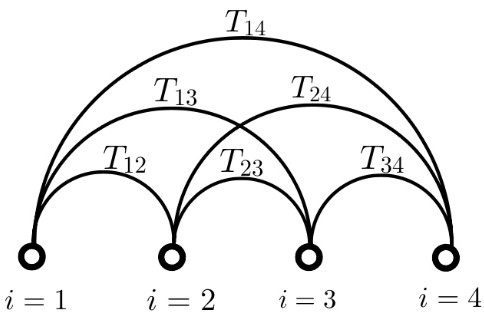}}_{\ket{0}}  \\
    +&\underbrace{\includegraphics[valign=c, width = 4cm]{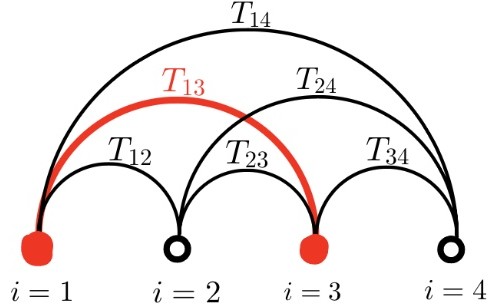}}_{ B_{13} \ \psi^\dagger_1 \psi^\dagger_3 \ket{0}} + \underbrace{\includegraphics[valign=c, width = 4cm]{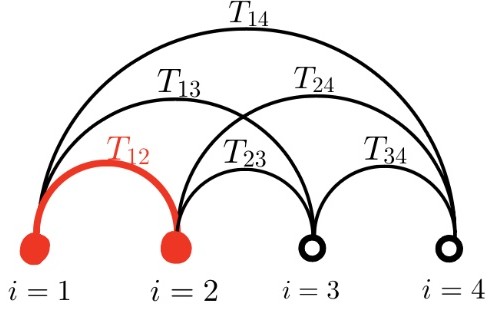}}_{ B_{12} \ \psi^\dagger_1 \psi^\dagger_2 \ket{0} } + \substack{\text{other diagrams with} \\ \text{one filled pair}} \nonumber \\
    +& \underbrace{\includegraphics[valign=c, width = 4cm]{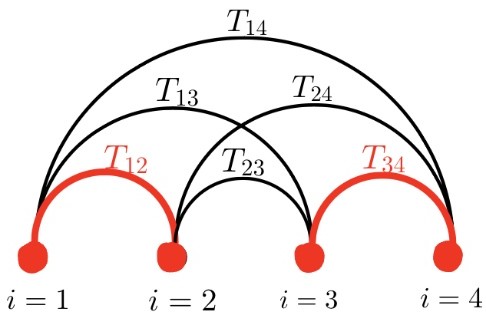}}_{B_{12} B_{34} \ \psi^\dagger_1 \psi^\dagger_2 \psi^\dagger_3 \psi^\dagger_4 } + \underbrace{\includegraphics[valign=c, width = 4cm]{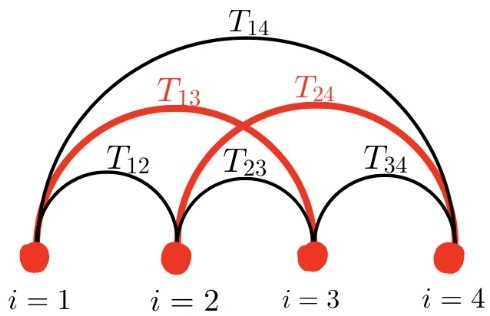}}_{B_{13} B_{24} \ \psi^\dagger_1 \psi^\dagger_3 \psi^\dagger_2 \psi^\dagger_4 } + \substack{\text{other diagrams with} \\ \text{two filled pairs}} \nonumber\\
\end{align}
which exactly reproduces \ref{eq:Gaussian_state}. In the same language the tensor product of two states simply amounts to $\ket{\text{adjecency matrices drawn next to each other}}$. Note that every node can only be colored once, since $\psi^\dagger \psi^\dagger = 0$. However, in the case of multiple fermionic modes per node (e.g. the different fermion colors in QCD) one can color each node and edge with all different colors. 

In the context of GGFPEPS it's often useful to consider the contraction of $\ket{B} \in \mathbf{H_{virt}}$ with $\ket{T} \in \mathbf{H_{phys}}\otimes\mathbf{H_{virt}} $ into $\ket{\psi} = \braket{B|T}$. Let us focus on the simple example where $\mathbf{H_{phys}} = \mathbf{H_{virt}} = \mathbb{C}^2$ with $\ket{B}$ is defined as :
\begin{align}
    \ket{B} = \del{1 + \psi^\dagger_1 \psi^\dagger_2}\ket{0} = \includegraphics[valign=c, width = 4cm]{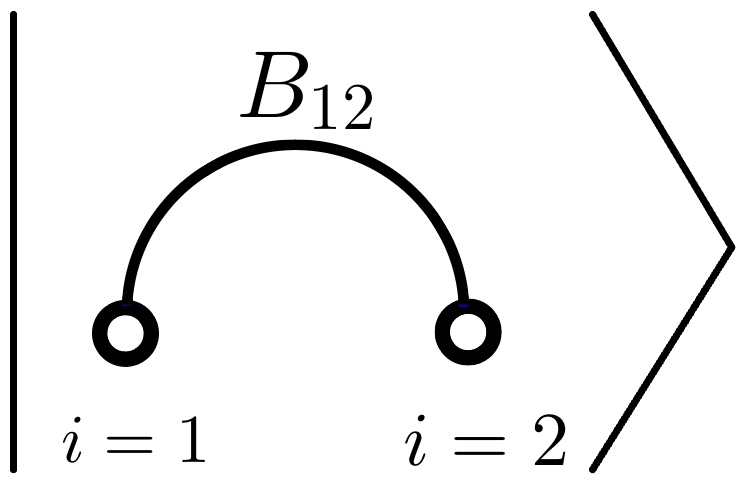}
\end{align}
and with this one can easily check that : 
\begin{align}
    \ket{\psi} = \braket{B|T} = \del{ 1+ B_{12}T_{12} } \ket{00} + \del{T_{34} + B_{12}T_{12}T_{34} + B_{12}T_{13}T_{24} + B_{12}T_{14}T_{23}  } \ket{11}
\end{align}
which we graphically represent as :
\begin{align}
    \braket{B|T} = \includegraphics[valign=c, width = 4cm]{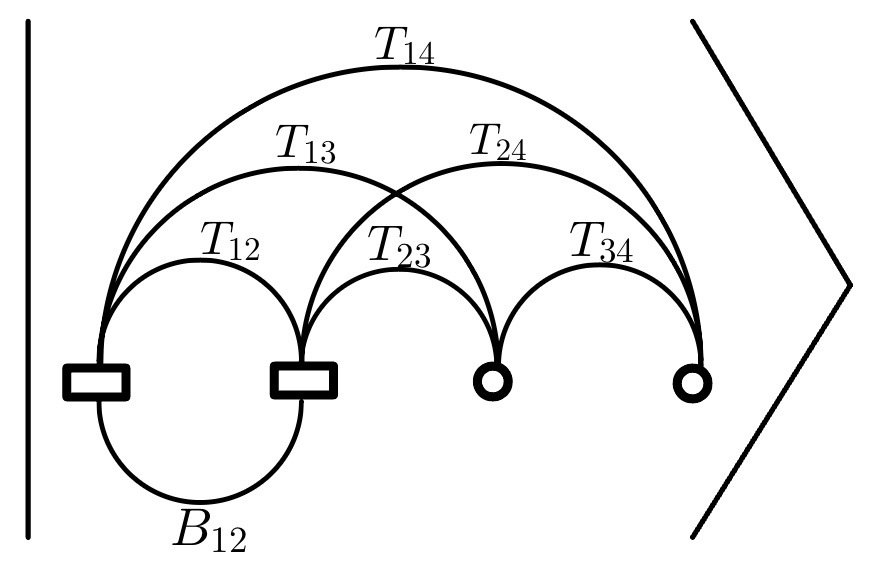}
\end{align}
supplemented with the rule that colored rectangular nodes (i.e. the ones that have been contracted over) must be connected to a colored edge from the bra and ket. Indeed, with these rules we get :
\begin{align}
    \ket{\psi} =& \includegraphics[valign=c, width = 4cm]{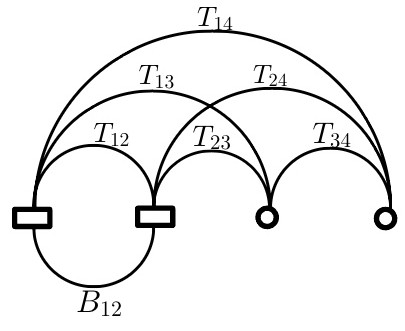} + \includegraphics[valign=c, width = 4cm]{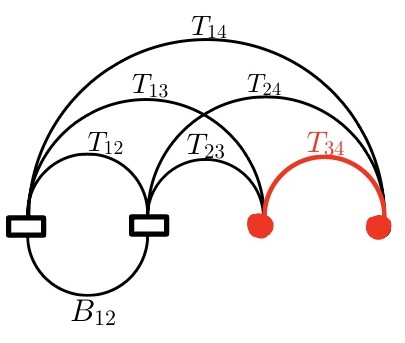}\\
    +&\includegraphics[valign=c, width = 4cm]{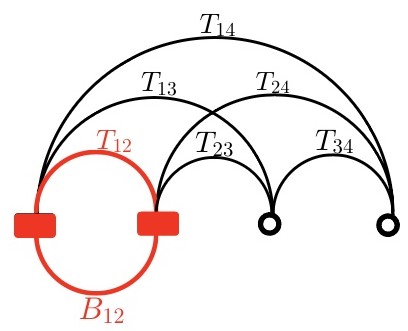} + \includegraphics[valign=c, width = 4cm]{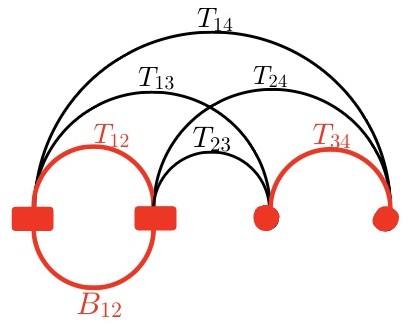} \nonumber \\
    +&\includegraphics[valign=c, width = 4cm]{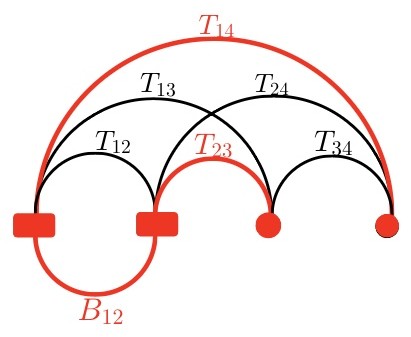} + \includegraphics[valign=c, width = 4cm]{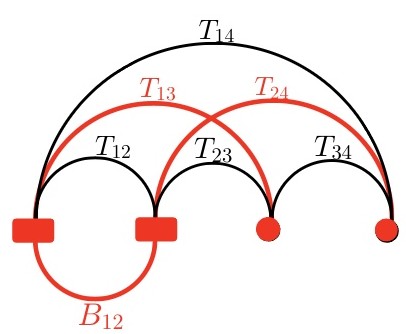} \nonumber
\end{align}
which exactly reproduces the analytical result.

\section{Fermionic coherent states and Grassmann variables}
\label{appendix:Fermionic coherent states and Grassmann variables}
When manipulating expressions containing fermionic creation and annihilation operators $\psi^\dagger_i$ and $\psi_i$, it is often convenient to use fermionic coherent states that are eigenvectors of the annihilation operator. 

More concrete, we would like to have states $\ket{\psi}$ so that $\hat \psi_i \ket{\psi} = \psi_i \ket{\psi} $. Since $\hat \psi_i \hat \psi_j = - \hat \psi_j \hat \psi_i$, consistency also requires that these eigenvalues satisfy $\hat \psi_i \psi_j = - \psi_j \hat  \psi_i $ and $\psi_i \psi_j = - \psi_j \psi_i $. The latter equation tells us that the eigenvalues must be Grassmann numbers. 

Below, we will review the basic properties of Grassmann numbers and use this to construct such states. We will focus on the case of a single fermion where $\psi^2 = 0$ allows us to truncate Taylor series for $f(x, \psi)$ exactly to first order in the Grassmann number $\psi$. The generalization to multiple fermions is straightforward, but some formulas will require some more care since $\psi_i \psi_j \neq 0 $ so that higher order terms in the Taylor expansion must be considered.

\subsection{Differentiation and integration with respect to Grassmann numbers}
With $\theta$ and $\eta$ denoting Grassmann numbers, the derivative is defined to satisfy : 
\begin{align}
    \partial_\theta (\theta) = 1 \text{ \hspace{1cm}and\hspace{1cm} }  \partial_\theta(1) = \partial_\theta(\eta) = 0   
\end{align}   
which is similar to that of normal numbers. However, to retain consistency with the anticommutation relations it is required that $\partial_\theta(\eta \theta) = - \partial_\theta(\theta \eta) = - \partial_\theta(\theta) \eta = - \eta$ holds. 

More surprisingly, the integral of a function of Grassmann variables is defined to satisfy 
\begin{align}
    \int d\theta f(\theta) = \partial_\theta f(\theta) 
\end{align}
which makes sense if we interpret the left-hand side as the Grassmannian equivalent of the improper integral $\int_{-\infty}^{\infty} dx f(x)$. Indeed, such improper integrals have the property they do no longer depend on $x$ and that $\int_{-\infty}^{\infty} dx f(x) = \int_{-\infty}^{\infty} dx f(x + a) \ \forall a$. It is easily checked that the above definition of the Grassmann integral also satisfies these properties.

On a more intuitive level this works because the improper integral $\int_{-\infty}^{\infty} dx f(x)$ is a map from a function $f(x)$ to a scalar. Since any function of Grassmann variables is at most linear in the Grassmann variables such a map from the function to the scalars can indeed be obtained by differentiation.  

\subsection{Properties of the fermionic coherent states}
Now that we are equipped with a basic understanding of Grassmann numbers it is relatively straightforward to build fermionic coherent states. Still for a single fermion we get :
\begin{align}
    \ket{\psi} = e^{- \psi \hat\psi^\dagger}\ket{0} \ \ .
\end{align}
is an eigenket of $\hat \psi$. Indeed, 
\begin{align}
    \hat \psi \ket{\psi} &= \hat \psi (1 - \psi \hat \psi^\dagger) \ket{0} = \psi \ket{0} = \psi (1 - \psi \hat \psi^\dagger) \ket{0} = \psi \ket{\psi} \ \ .
\end{align}

The overlap between these states is : 
\begin{align}
    \braket{\theta|\eta} &= \braket{0|e^{-\hat \psi \bar\theta} e^{-\eta \hat \psi^\dagger}|0} \\
    &= \braket{0|(1 -\hat \psi \bar\theta) (1-\eta \hat \psi^\dagger)|0} \\
    &= 1 + \bar\theta \eta \braket{0|\hat \psi   \hat \psi^\dagger|0} \\
    &= 1 + \bar\theta \eta  \\
    &= e^{\bar\theta \eta} 
\end{align}
and consequently the norm of a coherent state is $\mathcal{N}(\ket{\psi}) = \braket{\psi|\psi} = e^{\bar\psi \psi} = 1$.\

Furthermore, we can express the identity in terms of these coherent states as :
\begin{align}
    \one &= \int d\bar \theta d\theta e^{-\bar \theta \theta} \ket{\theta}\bra{\theta} \\
    &= \int d\bar \theta d\theta (1 -\bar \theta \theta)(1-\theta \hat \psi^\dagger)\ket{0}\bra{0}(1-\hat \psi \theta) \\
    &= \partial_{\bar \theta} \partial_\theta \del{ \ket{0}\bra{0} - \bar \theta \theta \ket{0}\bra{0} - \theta \ket{1}\bra{0} + \bar \theta \ket{0}\bra{1} + \theta \bar \theta \ket{1}\bra{1}  } \\
    &= \ket{0}\bra{0} + \ket{1}\bra{1} \ \  .
\end{align}

Finally, we will demonstrate that $\int d\omega \ e^{\omega(\eta - \theta)} = \delta(\eta - \theta)$. To do this, we first note that $\int d\omega e^{\omega(\eta - \theta)} = \eta - \theta$ so that :
\begin{align}
    \int d\eta \  \delta(\eta - \theta) f(\eta)  &= \int d \eta \ (\eta - \theta) (f_0 + f_1 \eta) \\
    &=  \partial_\eta \del{ \eta f_0 - \theta f_0 - \theta \eta f_1} \\
    &= f_0 + \theta f_1 = f(\theta) \ \ .
\end{align}

\section{Technical details of the pure gauge PEPS construction}
\label{appendix:technical details of the pure gauge PEPS construction}
To see that this is true we first use the fact that $e^x = \lim_{N \rightarrow \infty} \prod_{n\in 1:N} \del{1 + \frac{x}{N}}$. Truncating the product at some $N^{max}$ so that $x/N^{max} << 1$ then leads to 
\begin{align}
    \psi_I([\mathcal{G}]) &= \prod_{\substack{ \square \in plaquettes \\ t \\ n_\square }} \del{1 - \frac{\epsilon}{a n_\square}\frac{a^{D-3}}{g^2_\text{qft}} D^\square(\mathcal{G}_t)} \del{1 - \frac{\epsilon}{a n_\square}\frac{a^{D-3}}{g^2_\text{qft}} D^\square(\mathcal{G}^{-1}_t)} 
\end{align}
where we have dropped the indices $\bfn \, \bfei \, \bfej$ on $D^\square$ since these are already fixed by the plaquette index. To get the 2nd type of term we used $\bar D^\square(\mathcal{G}) = D^\square(\mathcal{G}^{-1})$. 

To simplify the notation we will now focus on a single term in the product and show that it can be rewritten as a simpler GGFPEPS i.e. we will show that 
\begin{align}
    1 - \frac{\epsilon}{a n_\square}\frac{a^{D-3}}{g^2_\text{qft}} D^\square(\mathcal{G}_t) = \braket{B|\mathcal{U}([\mathcal{G}])|A}_{\square}
\end{align}
where $\bra{B}_\square$, $\mathcal{U}([\mathcal{G}])_{\square}$ and $\ket{A}_\square$ are the one plaquette, time and flow direction versions of $\bra{B}_I$, $\mathcal{U}([\mathcal{G}])_I$ and $\ket{A}_I$ respectively. More specifically, these are defined by : 

\begin{align}
    \ket{A}_\square &= e^{\omega \del{a^\dagger b^\dagger+ c^\dagger d^\dagger f^\dagger e^\dagger + h^\dagger h^\dagger}}\ket{0} = \includegraphics[valign = c, width = 5cm]{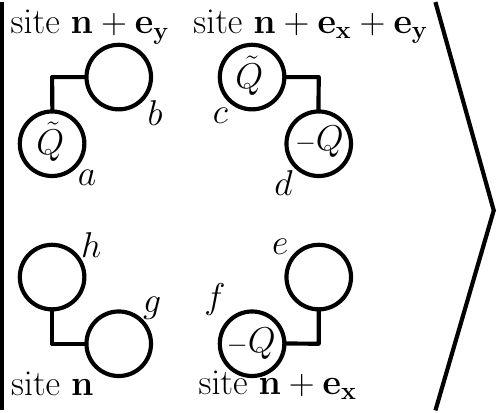}  \\
    \bra{B}_\square &= \bra{0} e^{ h a + b c + e d + g f} \\
    \mathcal{U}([\mathcal{G}]) &= e^{-i \phi_a(g_{bottom}) Q^a_f} e^{-i \phi_a(g_{right}) Q^a_d} e^{i \phi_a(g_{left}) \tilde Q^a_a} e^{i \phi_a(g_{top}) \tilde Q^a_c}   
\end{align}
where $\omega = \del{\frac{\epsilon}{a n_\square}\frac{a^{D-3}}{g_\text{qft}^2}}^{1/4}$, $g_{bottom} = g_{\bfn \, \bf + \bfex} $, $g_{right} = g_{\bfn + \bfex \, bfn + \bfex + \bfey}$, $g_{left} = g_{\bfn \, \bfn + \bfey}$ and $g_{top} = g_{\bfn + \bfey \, \bfn + \bfex + \bfey}$. Using the definition of the charges and their interplay with the annihilators (cfr. appendix \ref{appendix:Review on relevant group actions}) it is now easy to see that :

\begin{align}
    \bra{B}\mathcal{U}([\mathcal{G}])_\square &= \bra{0} e^{ h D(g_{left}) a + b D(g_{top}) c + e D^T(g^{-1}_{bottom}) d + g D^T(g^{-1}_{right}) f}
\end{align}
so that 
\begin{align}
    \bra{B}\mathcal{U}([\mathcal{G}])\ket{A}_\square &= 1 - \omega D_{\alpha \alpha'}(g_{left}) D_{\alpha' \beta'}(g_{top}) D_{\beta' \delta}(g^{-1}_{right}) D_{\delta \alpha}(g^{-1}_{bottom})   
\end{align}
which is exactly the desired term. From here the full correspondence between $\psi_I([\mathcal{G}])$ and $\braket{B|\mathcal{U}([\mathcal{G}])|A}_I$ follows by simply realizing that the latter is a direct product of terms like the one discussed above.

\section{Technical details of the fermionic PEPS construction}
\label{appendix:technical details of the fermionic PEPS construction}
To see that this is true we insert the definition of the relevant Hamiltonians into \ref{eq:first_def_fermionic_wavefunction} and use $\hat \varphi \ket{\varphi} = \varphi \ket{\varphi}$ and its conjugate $\bra{\varphi} \hat\varphi^\dagger = \bra{\varphi} \bar \varphi$ to get to
\begin{align}
    \ket{\psi([\mathcal{G}])}_{II} &= \int [D\varphi][D\bar \varphi ] \prod_{\bfn \, t} e^{-\bar \varphi_{\bfn \, t} \varphi_{\bfn \, t} } \prod_{\bfn \, t} e^{-\bar \varphi_{\bfn \,t} \varphi_{\bfn \, t+1}\del{1 + \epsilon m + (-1)^\bfn \epsilon \mu }} \\ 
    &\times \prod_{\braket{\bfn \bfm} \, t} e^{-i \frac{\epsilon}{a}\alpha^*_{\bfn \bfm} \bar \varphi_{\bfn \,t} \Dph^*_{\bfn \bfm}(g_t)\bar \varphi_{\bfm \, t} }  e^{-i \frac{\epsilon}{a}\alpha_{\bfn \bfm} \varphi_{\bfn \, t+1} \Dph_{\bfn \bfm}(g_{t+1})\varphi_{\bfm \, t+1} } \nonumber 
\end{align} 
which is now purely in terms of Grassmann variables. Next, we introduce new Grassmann fields $\varphi^C_{\bfn \, \bfei \, t}$ and $\varphi^D_{\bfn \, \bfei \, t}$ for each site $\bfn$, direction $\bfei$ and time $t$ to rewrite this as : 
\begin{align}
    \ket{\psi([\mathcal{G}])}_{II} &= \int [D\varphi^C][D\bar \varphi^C] \prod_{\bfn \, \bfei 
    , t}e^{-\bar \varphi^C_{\bfn \, \bfei \,t}\varphi^C_{\bfn \, \bfei \, t} }\int [D\varphi^D][D\bar \varphi^D] \prod_{\bfn \, \bfei \, t}e^{-\bar \varphi^D_{\bfn \, \bfei \, t+1}\varphi^D_{\bfn \, \bfei \,t+1} }  \\
    &\times \prod_{\braket{\bfn \bfm} \, t} e^{-i \alpha^*_{\bfn \bfm} \varphi^C_{\bfn \, \bfm - \bfn \, t} \Dph^*_{\bfn \bfm}(g_t) \varphi^C_{\bfm \, \bfn-\bfm \, t} } \prod_{\braket{\bfn \bfm}\, t} e^{-i \alpha_{\bfn \bfm} \varphi^D_{\bfn \, \bfm - \bfn \, t+1} \Dph_{\bfn \bfm}(g_{t+1}) \varphi^D_{\bfm \, \bfn-\bfm \,t+1} }  \nonumber \\
    &\hspace{1cm}\times \prod_\bfn \int d\varphi_{\bfn \, t=0}d\bar\varphi_{\bfn \, t=0} \ket{\varphi_{\bfn \, t=0}} e^{-\bar\varphi_{\bfn \, t=0}\varphi_{\bfn \, t=0}}e^{-\bar\varphi_{\bfn \, t=0}\varphi_{\bfn \, t=1}r_\bfn} \nonumber \\
    &\hspace{1cm}\times \prod_{t = 1}^{N-1} \int d\varphi_{\bfn \, t}d\bar\varphi_{\bfn \, t} e^{-\bar\varphi_{\bfn \, t}\varphi_{\bfn \, t}}e^{-\bar\varphi_{\bfn \, t}\varphi_{\bfn \, t+1}r_\bfn} e^{\sqrt{\frac{\epsilon}{a}}\sum_{\bfei}\del{ \bar \varphi^C_{\bfn \, \bfei \, t}\bar \varphi_{\bfn \, t} + \bar \varphi^D_{\bfn \, \bfei \, t} \varphi_{\bfn \, t}} } \nonumber \ \ .
\end{align}

To see that this is true, it suffices to note that $\int D\bar \varphi^D e^{- \sum_\bfei \bar\varphi^D_{\bfn\, \bfei \, t+1} \del{ \varphi^D_{\bfn \, \bfei \, t+1} - \sqrt{\frac{\epsilon}{a}}\varphi_{\bfn \, t} } } = \prod_\bfei \delta( \varphi^D_{\bfn \, \bfei \, t+1} - \sqrt{\frac{\epsilon}{a}}\varphi_{\bfn, t+1} )$ and $\int D\varphi^C e^{- \sum_\bfei \varphi^C_{\bfn \, \bfei \, t} \del{ \varphi^C_{\bfn \, \bfei \, t} - \sqrt{\frac{\epsilon}{a}}\bar\varphi_{\bfn \, t} } } = \prod_\bfei \delta(\varphi^C_{\bfn \, \bfei \, t} - \sqrt{\frac{\epsilon}{a}}\bar\varphi_{\bfn \, t}).  $
where the $\varphi_{\bfn \, t}$ on different sites now no longer couple. Due to the insertion of these new Grassmann fields the remaining $\varphi_{\bfn \, t}$ integrals are Gaussian so that they can be solved analytically, this is done in reference \cite{Emonts_Fermionic_Gaussian_projected_entangled_pair_states_in_3+1D}. Applying their result gives us : 
\begin{align}
    \ket{\psi([\mathcal{G}])}_{II} &= \int [D\varphi^C][D\bar \varphi^C] \prod_{\bfn \, \bfei 
    , t}e^{-\bar \varphi^C_{\bfn \, \bfei \,t}\varphi^C_{\bfn \, \bfei \, t} }\int [D\varphi^D][D\bar \varphi^D] \prod_{\bfn \, \bfei \, t}e^{-\bar \varphi^D_{\bfn \, \bfei \, t+1}\varphi^D_{\bfn \, \bfei \,t+1} }  \\
    &\times \prod_{\braket{\bfn \bfm} \, t} e^{-i \alpha^*_{\bfn \bfm} \varphi^C_{\bfn \, \bfm - \bfn \, t} \Dph^*_{\bfn \bfm}(g_t) \varphi^C_{\bfm \, \bfn-\bfm \, t} } \prod_{\braket{\bfn \bfm}\, t} e^{-i \alpha_{\bfn \bfm} \varphi^D_{\bfn \, \bfm - \bfn \, t+1} \Dph_{\bfn \bfm}(g_{t+1}) \varphi^D_{\bfm \, \bfn-\bfm \,t+1} }  \nonumber \\
    &\hspace{0.7cm}\times\int d\varphi_{\bfn \, t=0}d\bar \varphi_{\bfn \, t=0}  e^{-\bar\varphi_{\bfn \, t=0}\varphi_{\bfn \, t=0}}\ket{\varphi_{\bfn \, t=0}} \nonumber \\
    &\hspace{0.7cm}\times \exp\del{-\bar\varphi_{\bfn \, t=0} \sum_{\substack{t=0 \\ \bfei}}^{N-1} \del{r_\bfn \sqrt{\frac{\epsilon}{a}}}^t  \bar \varphi^C_{\bfn \, \bfei \, t}  + \sum_{\substack{0 \leqslant t \leqslant s \leqslant N-1 \\ \bfei \, \bfej }} \del{r_\bfn \frac{\epsilon}{a}}^{t-s} \bar \varphi^C_{\bfn \, \bfei \, t} \bar \varphi^D_{\bfn \, \bfej \, s}} \nonumber 
\end{align}
which is still just a function of Grassmann variables. To recast this as an expression in terms of Hilbert spaces we use $ \braket{0^C 0^D | \varphi^C \varphi^D} \braket{\varphi^C \varphi^D | 0^C 0^D} = 1 $, $\braket{\varphi|0}_{II} = 1$ and the resolution of the identity for coherent states. This leads us to $\ket{\psi([\mathcal{G}])}_{II} = \braket{B(\mathcal{G})|A}_{II}$ with $\ket{A}$ as prescribed above and :
\begin{align}
    \bra{B([\mathcal{G}])} = \bra{0}_{II} \prod_{\braket{\bfn \bfm} \, t} &\exp\del{-i \alpha^*_{\bfn \bfm} \varphi^C_{\bfn \, \bfm - \bfn \, t} \Dph^*_{\bfn \bfm}(g_t) \varphi^C_{\bfm \, \bfn-\bfm \, t}}  \\
    &\hspace{1.5cm}\times\exp\del{-i \alpha_{\bfn \bfm} \varphi^D_{\bfn \, \bfm - \bfn \, t+1} \Dph_{\bfn \bfm}(g_{t+1}) \varphi^D_{\bfm \, \bfn-\bfm \,t+1} } \nonumber \ \ .
\end{align}

It is then easy to see that $\bra{B([\mathcal{G}])} = \bra{B} \mathcal{U}([\mathcal{G}])$ since the gauging operation has been defined so that 
\begin{align}
    \mathcal{U}^\dagger([\mathcal{G}]) \, \psi^C_{\bfm \, \bfn-\bfm \,t}\, \mathcal{U}([\mathcal{G}]) &= \Dph^*_{\bfn \bfm}(g_t) \psi^C_{\bfm \, \bfn-\bfm \, t} \\ 
    \mathcal{U}^\dagger([\mathcal{G}]) \, \psi^D_{\bfm \, \bfn-\bfm \, t} \, \mathcal{U}([\mathcal{G}]) &= \Dph_{\bfn \bfm}(g_t) \psi^D_{\bfm \, \bfn-\bfm \, t} 
\end{align}
and $\bra{0^C}\bra{0^D}\mathcal{U}^\dagger([\mathcal{G}]) = \bra{0^C}\bra{0^D}$ since the vacuum carries no charge.

\bibliography{biblio}

\end{document}